\documentclass[conference]{IEEEtran}
\IEEEoverridecommandlockouts
\usepackage{cite}
\usepackage{ifthen}
\usepackage{tabularx}
\usepackage{graphicx}
\usepackage{arydshln}
\usepackage{changepage} % This is for \adjustwidth (used for tables)
\usepackage{amsmath}
\usepackage{amssymb}
\usepackage{wasysym}
\usepackage{nccmath}
\usepackage{caption}
\usepackage{listings}
\usepackage{enumitem} 
\usepackage{xspace} 
\usepackage{xcolor}

\usepackage{url}
\usepackage[justification=centering]{caption}

\lstdefinelanguage{mafia}{
  classoffset = 1,
  morekeywords = {clock, window, async, match, random, Random, HashMap, Counter, Sketch, BloomFilter, Timestamp, timestamp, counter, tag, sample, duplicate, collect, bloomfilter, sketch, Key, key, alg, type, nhash, hash, size, width, store, all, any, add, sub, rem, min, max, set, reset, update, test, insert, count, (, ), <, >},
  keywordstyle={\itshape\bfseries},
  classoffset = 2,
  morekeywords = {sum, min},
  keywordstyle = \itshape,
  morecomment = [l]{//},
  morecomment = [s]{/*}{*/},
  morecomment = [s]{/**}{*/},
  commentstyle = \color{gray},
  morestring = [b]",
  morestring = [b]'
}
\newboolean{showcomments}
\setboolean{showcomments}{true}
\ifthenelse{\boolean{showcomments}}
{ }

\newcommand{\smartparagraph}[1]{\noindent{\bf #1}\ }
\newcommand{\eg}{e.g., }
\newcommand{\ie}{i.e., }

\newcommand{\mafia}{MAFIA\xspace}

\newcommand{\mmentint}{{\sf Window}\xspace}
\newcommand{\match}{{\sf Match}\xspace}
\newcommand{\ts}{{\sf Timestamp}\xspace}
\newcommand{\counter}{{\sf Counter}\xspace}
\newcommand{\tagg}{{\sf Tag}\xspace}
\newcommand{\sample}{{\sf Sample}\xspace}

\newcommand{\bloom}{{\sf BloomFilter}\xspace}
\newcommand{\sketch}{{\sf Sketch}\xspace}

\newcommand{\seqlsym}{\textbf{$\gg$}}
\newcommand{\parlsym}{\textbf{$+$}}
\newcolumntype{L}[1]{>{\raggedright\let\newline\\\arraybackslash\hspace{0pt}}m{#1}}
\newcolumntype{C}[1]{>{\centering\let\newline\\\arraybackslash\hspace{0pt}}m{#1}}
\newcolumntype{R}[1]{>{\raggedleft\let\newline\\\arraybackslash\hspace{0pt}}m{#1}}

\newcommand{\onecolmn}[3]
{
	\begin{center}
    	\begin{tabular}{#1{#2}}
    		#3\\
    	\end{tabular}
    \end{center}
}

\newcommand{\twocolmn}[6]
{
	\begin{center}
    	\begin{tabular}{@{\extracolsep{\fill}}C{#3}|C{#5}}
    		\multicolumn{2}{C{#1}}{#2}\\#4&#6\\
    	\end{tabular}
    \end{center}
}

\begin{document}
%\title{EasyMeas: Making The Specification of Network Measurement Tasks  Great Again}
\title{Measurements As First-class Artifacts \\ \large{\textsl{Extended Version}}}
%\titlenote{Produces the permission block, and copyright information}

%\author{Anonymous}

\author{Paolo Laffranchini$^{\star\diamond\Box}$\thanks{$^{\Box}$Work done in part while visiting at KAUST.} \quad
Luis Rodrigues$^\star$ \quad
Marco Canini$^\dag$ \quad
Balachander Krishnamurthy$^\ddag$ \\
$^\star$ INESC-ID, IST, U. Lisboa \quad
$^\diamond$ Universit\'e catholique de Louvain \quad
$^\dag$ KAUST \quad
$^\ddag$ AT\&T Labs -- Research
}

% \author{Paolo Laffranchini}\authornote{Part of this work has been done during an exchange period in KAUST}
% \affiliation{%
%   \institution{INESC-ID, Instituto Superior T\'ecnico}
%   %\city{Lisbon} 
%   %\state{Portugal} 
% }
% \email{paolo.laffranchini@tecnico.ulisboa.pt}

% \author{Luis Rodrigues}
% \affiliation{%
%   \institution{INESC-ID, Instituto Superior T\'ecnico}
%   %\city{Lisbon} 
%   %\country{Portugal}
% }
% \email{ler@tecnico.ulisboa.pt}

% \author{Marco Canini}
% \affiliation{%
%   \institution{KAUST}
%   %\city{Thuwal} 
%   %\state{Saudi Arabia} 
% }
% \email{marco@kaust.edu.sa}

% \author{Balachander Krishnamurthy}
% \affiliation{%
%   \institution{AT\&T Labs - Research}
%   %\city{New York}
%   %\country{USA}
% }
% \email{bala@research.att.com}

%\newcommand{\ourlang}{NMSL\xspace}
%\newcommand{\ourlanguage}{NMSL\xspace}
%\newcommand{\ourlanguage}{\mafia\xspace}
\newcommand{\ourlang}{\mafia}

\maketitle

\setlength{\textfloatsep}{3pt}
\setlength{\abovecaptionskip}{3pt}

\begin{abstract}

The emergence of programmable switches has sparked a significant amount of work on new techniques to perform more powerful measurement tasks, for instance, to obtain fine-grained traffic and performance statistics. Previous work has focused on the efficiency of these measurements alone and has neglected flexibility, resulting in solutions that are hard to  reuse or repurpose  and that often overlap in functionality or goals.

In this paper, we propose the use of a set of reusable primitive building blocks that  can be composed  to express measurement tasks in a concise and simple way. We describe the rationale for the design of our primitives, that we have named MAFIA (Measurements As FIrst-class Artifacts), and using several examples we illustrate  how they can be combined to realize a comprehensive range of network measurement tasks. Writing \ourlang code does not require expert knowledge of low-level switch architecture details. Using a prototype implementation of \ourlang, we demonstrate the applicability of our approach and show that the use of our primitives results in compiled code that is comparable in size and resource usage with manually written specialized P4 code, and can be run in current hardware. 
\end{abstract}

%\begin{CCSXML}
%\end{CCSXML}

%\ccsdesc[500]{Computer systems organization~Embedded systems}

%\keywords{SDN, Network Measurement, Programmable Networks}

\section{Introduction}
\label{sec:intro}

Historically, network measurement's evolution paralleled the growth of
the Internet but at a much slower pace. SNMP, {\it ping}, and {\it
traceroute} constituted the bulk of measurement-related aids for a
long time. The introduction of SDN has led to significant work on
various aspects of programmable network infrastructures. An SDN {\it
controller} can dynamically install and modify switch rules, enforce
high-level operator policies and gather statistics. Starting from the
original white paper\cite{sdn-white-paper} various aspects of SDN (and
particularly OpenFlow\cite{OpenFlow}) have been examined in depth. Unfortunately,
measurement, a well-understood requirement for the Internet, with a
long body of developed work for over two decades, appears to have been
an afterthought in SDN's development.  In fact, \cite{sdn-white-paper}
mentions {\it security} a dozen times (rightfully so) but the words
measurement or metrics do not appear in it.

Given measurement's importance in network operation and management,
there has been a flurry of work on exploiting SDN features and
programmable switches to perform more powerful measurement tasks. 
%A variety of new measurement ideas have been discussed in the literature
%and some ideas have been deployed in the field.  
Beyond OpenFlow, proposals like OpenState~\cite{OpenState} and switch
programmability as in P4~\cite{P4} have enabled richer, customizable
in-network processing that can implement measurements for fine-grained
traffic and network performance statistics~\cite{Marple, turboflow,
StarFlow}. Most of the recent work in this area focuses on 
efficiently mapping measurement tasks on programmable forwarding
elements.  Efficiency is key as current programmable switch chips have
limited computational and memory
resources~\cite{ForwardingMetamorphosis, Marple, StarFlow, DAIET}.

An important requirement that has not been addressed in prior work is
flexibility and extensibility in supporting a variety of measurement
tasks; instead we have ad-hoc solutions proposed for specific
measurements. In spite of advances in programmable data planes, it is not possible without significant effort to combine,
reuse or repurpose existing solutions although they may partly overlap
in functionality or goals.

We instead argue for supporting flexible measurement through a set of
reusable building blocks (\emph{primitives}) that take advantage of novel
features of programmable forwarding elements and span most of the
commonly performed measurement tasks. We identify a set of such
primitives that network operators can use to
express measurement tasks in a concise and simple way. Further, they
are reusable as complex tasks can be expressed by composing a few
calls to a subset of our measurement primitives.

We define our approach as Measurements As FIrst-class Artifacts, or \ourlang for short.
Concretely, we instantiate our ideas as an API that provides an abstraction over measurement primitives that execute at line rate in the data plane. We remark that our primary target is network operators, who are not proficient data plane programmers, yet they desire to quickly address performance-, security- and troubleshooting-related measurement needs. As such, our goal is not satisfied by and is orthogonal to data plane programming languages like P4. These technologies are an enabler for \ourlang but remain fundamentally lower-level approaches.

%We implement and the evaluation shows that the overheads are low.

% In this paper we advocate \RM{an alternative} \pl{a principled]} approach \mcnote{alternative to what?} that aims at making the task of specifying network measurements tasks simpler, more \pl{concise} \RM{elegant} \mcnote{Not sure elegant is a driving consideration. Conciseness might be.}, \pl{via a set of common abstraction which can be efficiently reused for different purposes} \RM{and more reusable}. \mcnote{More than what?} In our approach, complex measurement tasks can be easily expressed by declaring sequential and parallel compositions of few calls to a selected set of measurement primitives. 
% \mcnote{Abstract and sec 2 refer to this as a programming model. If that is the case, it should be made explicit here too. Do we say why an ``alternative'' programming model is needed? Are we suggesting that current programming models are not good? Part of the earlier motivation was rather concerned with the fact that earlier measurement efforts were ad hoc and could not be readily reused. Is that still part of the problem, or do we pitch P4++ here?}
%In this paper we describe the rationale for the design of our primitives and illustrate, using several examples, how they can be combined to specify a comprehensive  range of network measurement tasks. Using a prototype implementation of a \ourlanguage compiler, we show that it is possible to generate runnable code that compares in complexity and efficiency to hand-written P4 code.

Our work is informed by the large number of legacy measurements that
have been carried out routinely in large and small networks as well as
new ones in the SDN milieu. We identify the primitives for measurement
on the basis of their {\it breadth} of applicability and the ability
for {\it maximal reuse} (i.e., a good implementation can yield rich
dividends in a broad set of contexts). 
We are driven by four key considerations inherent in measurement~\cite{candk}: 
 {\it what}, {\it where}, {\it when}, and {\it how}. 
% Furthermore, we are also concerned with cost, given that measurement tasks 
% should not impair but, instead, support better forwarding (forwarding policies 
% benefit from the information collected by measurement). 
We validate our idea by showing that several key known SDN
measurements and some new ones can be built by composing our
abstractions. Our primitives can be used to answer questions ranging from
network-wide traffic characteristics (e.g., flow size distributions,
identifying heavy hitters~\cite{openTM, HHH, OpenSketch, UnivMon}, to
fine-grained monitoring of properties of flows and switches
(throughput, latency, loss,
etc.)~\cite{OpenNetmon, opensample, FlowRadar, LossRadar, DAPPER}, to
verification (traffic behavior matching operator's
intent)~\cite{SDNTraceroute}, to debugging (\eg troubleshooting root
causes of performance problems or switch/controller
misbehavior)~\cite{EverFlow,netsight}, and various security aspects
(\eg anomalies, DDoS, malicious
activity)~\cite{OpenSketch, UnivMon}.

%%%%%%%%%%%%%%%%%%%%%%%%%%%%%%%%%%%%%%%%%%%%%%%%%%%%%%%%%%%%%%%%%%%%%%%%%%%%%%%%%%%%%%%%%%%%%%%%%%%%%%%%%%%%%%%%%%%% 
%%%%%%%%%%%%%%%%%%%%%%%%%%%%%%%%%%%%%%%%%%%%%%%%%%%%%%%%%%%%%%%%%%%%%%%%%%%%%%%%%%%%%%%%%%%%%%%%%%%%%%%%%%%%%%%%%%%% 
%%%%%%%%%%%%%%%%%%%%%%%%%%%%%%%%%%%%%%%%%%%%%%%%%%%%%%%%%%%%%%%%%%%%%%%%%%%%%%%%%%%%%%%%%%%%%%%%%%%%%%%%%%%%%%%%%%%% 
											% TABLE 1 %
\begin{table}[t!]
\scriptsize
\begin{adjustwidth}{-0.35cm}{}
%\begin{center}
\begin{tabular}{m{0.8em} | m{6em} : m{7em} : m{6em} : m{9em}|}

%%%%%%%%%%%%%%%%%%%%%%%%%%%%%%%%%%%%%%%%%%%%%%%%%%%%%%%%%%%%%%%%%%%%%%%%%%%%%%%%%%%%%%%%%%%%%%%%%%%%%%%%%%%%%%%%%%%%
\cline{2-5}
		\multicolumn{1}{c|}{} & 
        \onecolmn{@{\extracolsep{\fill}}C}{6em}{\textbf{WHAT}} &
        \twocolmn{6em}{\textbf{WHERE}}{3em}{\textit{Legacy}}{2em}{\textit{SDN}} &
        \twocolmn{5em}{\textbf{WHEN}}{2.9em}{\textit{Legacy}}{2em}{\textit{SDN}} &
        \twocolmn{8em}{\textbf{HOW}}{4em}{\textit{Legacy}}{4em}{\textit{SDN}} 

%%%%%%%%%%%%%%%%%%%%%%%%%%%%%%%%%%%%%%%%%%%%%%%%%%%%%%%%%%%%%%%%%%%%%%%%%%%%%%%%%%%%%%%%%%%%%%%%%%%%%%%%%%%%%%%%%%%% 

\\\cline{2-5}\cline{2-5}
		\rotatebox[origin=c]{90}{\textbf{Traffic Eng.}} &
        \onecolmn{@{\extracolsep{\fill}}C}{6em}{Traffic Matrix; Flow size distribution;\\ Changes; \\ Anomalies; Heavy Hitters.} &
        \twocolmn{6em}{Online; \newline Network Wide}{3em}{Server}{3em}{Ctrl \newline Plane} &
        \onecolmn{@{\extracolsep{\fill}}C}{6em}{Always}  &
        \twocolmn{9em}{}{4em}{SNMP; NetFlow; sFlow}{4em}{Counters; Samples; Sketches}

%%%%%%%%%%%%%%%%%%%%%%%%%%%%%%%%%%%%%%%%%%%%%%%%%%%%%%%%%%%%%%%%%%%%%%%%%%%%%%%%%%%%%%%%%%%%%%%%%%%%%%%%%%%%%%%%%%%% 

\\\cline{2-5}
        \rotatebox[origin=c]{90}{\textbf{Performance}} &
        \onecolmn{@{\extracolsep{\fill}}C}{6em}{Volume; \\ Throughput; \\ Latency/Jitter; \\ Queue Length; \\ Packet Loss.} &
        \onecolmn{C}{6em}{Online; \newline Network Wide}  &
        \onecolmn{@{\extracolsep{\fill}}C}{6em}{Always; \newline QoS/SLA \newline Violation} &
        \twocolmn{9em}{}{4em}{SNMP; NetFlow; sFlow}{4em}{Counters; Probes; Samples; Bloom filters; Sketches}

%%%%%%%%%%%%%%%%%%%%%%%%%%%%%%%%%%%%%%%%%%%%%%%%%%%%%%%%%%%%%%%%%%%%%%%%%%%%%%%%%%%%%%%%%%%%%%%%%%%%%%%%%%%%%%%%%%%% 

\\\cline{2-5}
		\rotatebox[origin=c]{90}{\textbf{Verification}} &
        \onecolmn{@{\extracolsep{\fill}}C}{6em}{Network Invariants; \newline Routing Policies. } & %\newline NFV Chaining; \newline Control/Data Plane Behavior} &
        \twocolmn{6em}{Offline; \newline Network Wide}{3em}{--}{3em}{Flow Tables; Packet} &
        \twocolmn{5em}{Prior \newline Deployment}{2.5em}{--}{2.5em}{Runtime} &
        \twocolmn{8em}{Config. Analysis}{4em}{-}{4em}{Tags;}

%%%%%%%%%%%%%%%%%%%%%%%%%%%%%%%%%%%%%%%%%%%%%%%%%%%%%%%%%%%%%%%%%%%%%%%%%%%%%%%%%%%%%%%%%%%%%%%%%%%%%%%%%%%%%%%%%%%% 

\\\cline{2-5} 
		\rotatebox[origin=c]{90}{\textbf{Troubleshooting}} & 
        \onecolmn{@{\extracolsep{\fill}}C}{6em}{Packet Loss; \\ Network Invariants; \newline Routing Policies.} & % \\ Control/Data Plane Behavior } &
        \twocolmn{6em}{Online; \newline Network Wide}{3em}{Switch;\newline Router}{3em}{Ctrl Plane; Flow Tables; Packet} &
        \onecolmn{@{\extracolsep{\fill}}C}{6em}{On-demand (after issue notification)} &
        \twocolmn{9em}{}{4em}{ping; tracert; iperf; \newline SNMP; \newline Log Analysis}{4em}{Counters; \newline Tags; \newline Bloom filters;}\\\cline{2-5} 

%%%%%%%%%%%%%%%%%%%%%%%%%%%%%%%%%%%%%%%%%%%%%%%%%%%%%%%%%%%%%%%%%%%%%%%%%%%%%%%%%%%%%%%%%%%%%%%%%%%%%%%%%%%%%%%%%%%% 

			\rotatebox[origin=c]{90}{\textbf{Security}} &  
			\onecolmn{@{\extracolsep{\fill}}C}{6em}{DDoS; \newline Superspreaders; \newline Intrusion Detection.} &
            \onecolmn{@{\extracolsep{\fill}}C}{7em}{Network Wide} &
            \onecolmn{@{\extracolsep{\fill}}C}{6em}{Always} &
            \twocolmn{8em}{IDS}{4em}{SNMP; NetFlow; sFlow}{4em}{Counters; Sketches; Samples }\\\cline{2-5}
			
%%%%%%%%%%%%%%%%%%%%%%%%%%%%%%%%%%%%%%%%%%%%%%%%%%%%%%%%%%%%%%%%%%%%%%%%%%%%%%%%%%%%%%%%%%%%%%%%%%%%%%%%%%%%%%%%%%%%

\end{tabular}
%\end{center}
\end{adjustwidth}
%  \vspace{2mm}  
  \caption{Measurements scenarios: Legacy vs SDN.}
  %\caption*{\footnotesize \textit{\textbf{TE}}: Traffic Engineering; \textit{\textbf{PM}}: Performance Monitoring;\textit{\textbf{V}}: Verification;\textit{\textbf{TS}}: Toubleshooting;\textit{\textbf{S}}: Security;}
  \label{tab:table-1}
\end{table}
											% TABLE 1 %
%%%%%%%%%%%%%%%%%%%%%%%%%%%%%%%%%%%%%%%%%%%%%%%%%%%%%%%%%%%%%%%%%%%%%%%%%%%%%%%%%%%%%%%%%%%%%%%%%%%%%%%%%%%%%%%%%%%% %%%%%%%%%%%%%%%%%%%%%%%%%%%%%%%%%%%%%%%%%%%%%%%%%%%%%%%%%%%%%%%%%%%%%%%%%%%%%%%%%%%%%%%%%%%%%%%%%%%%%%%%%%%%%%%%%%%% %%%%%%%%%%%%%%%%%%%%%%%%%%%%%%%%%%%%%%%%%%%%%%%%%%%%%%%%%%%%%%%%%%%%%%%%%%%%%%%%%%%%%%%%%%%%%%%%%%%%%%%%%%%%%%%%%%%%

We contribute the following:
\textit{i)} We identify a set of programmable
and reusable \textsl{primitives} that can be supported by switches to
realize flexible measurement tasks;
\textit{ii)} We show how our
primitives can be composed and applied to a wide variety of
measurements due to their orthogonality;
\textit{iii)} We develop a
\ourlang prototype that compiles measurements expressed through our API into equivalent P4 codes that can execute in current P4-compliant programmable switches. We show that our abstractions reduce development effort of measurement tasks while the resulting P4 code is, in size, only marginally larger than  the hand-written version, and can be mapped to hardware with a modest use of resources.
\ourlang is released at \url{https://github.com/paololaff/mafia-sdn}.

\section{On Measurement Primitives}
\label{sec:dsl}

A core tenet of our work is that many of the common network
measurement tasks can be expressed by \emph{composing primitives} that
can be supported by current and future programmable forwarding
elements.  But what is a good primitive? Functions that are
routinely applicable for a range of measurement needs is a candidate
given its potential for reuse. Functions also need to be composable to
express more complex tasks. They should be sufficiently
low-level to be broadly applicable but sufficiently high-level to
reduce effort.

However, modern data rates of high-speed networks impose stringent
per-packet processing requirements. Thus, primitives should have low processing and state complexity so they can be
implemented in programmable forwarding elements.

Finally, the primitives need not be novel; instead, we seek to ground
our choice on functionalities that have proven themselves useful in
various contexts. We survey prior work, studying a range of
measurement scenarios in traditional networking and in SDN
environments before describing the set of our primitives.

%\label{sec:scenarios}
%\footnote{Due to space limit, we omit citations for all but the most relevant works.} 

\noindent\noindent\textit{Measurement Scenarios:} We performed an analysis of the extensive related work in network
measurements. We examined five key categories that
have historically dominated work in this area: traffic
engineering\cite{openTM,HHH,DREAM,OpenSketch,SCREAM,UnivMon},
performance
monitoring\cite{OpenNetmon,opensample,FlowRadar,LossRadar,DAPPER,Marple,StarFlow},
verification\cite{INT,VeriDP,SDNTraceroute}, troubleshooting \cite{netsight, EverFlow,StarFlow} and
security\cite{OpenSketch,SCREAM,UnivMon}.
%\mcnote{Didn't we have more citations? We also used to have citations in the table. It doesn't seem that exhaustive the current citation list.}  \mcnote{Does the above list of citations include all that needs to be cited? Where are other citations that are included elsewhere in the paper, like INT, TPP, turboflow, StarFlow, MARPLE, etc.?}

While in legacy networks, SNMP, NetFlow and sFlow were still the tools 
(despite being ineffective in enabling visibility into the details of individual flows), 
in SDN context, packet and byte counters available from OpenFlow matching 
rules help\cite{OpenNetmon} in calculating throughput, port/link utilizations 
and packet loss with their exact information. Polling frequency is traded off 
against computational cost. Sampling is an option\cite{opensample} if 
protocol-specific information like sequence numbers are available to correct 
the estimation phase. Novel approaches have also proposed compact data 
structures and algorithms based on Bloom filters to monitor the number of 
packets for each flow\cite{FlowRadar} as well as packet 
losses\cite{FlowRadar,LossRadar}. Sketches can as well be used for these 
measurement scenarios to provide approximate counters for a group of flows. 
Programmable forwarding elements are also enabling stateful tracking of flows directly in the switches\cite{DAPPER}.

Table~\ref{tab:table-1} summarizes the main measurement categories and captures the
differences between traditional and SDN environments. We identified
common factors exploited for measurement-related problems. The
mechanisms used in previous measurements (shown in the last two
columns) are particularly relevant for our work, as they helped us
identify a small set of key primitives usable in a broad range of use
cases. We build on this small set of building blocks to provide
flexible, programmable measurements for most of the known measurement
tasks.

%\label{sec:apirationale}
\vspace{2mm}
\noindent\noindent\textit{Selecting the Primitives:} Examining the various techniques that have been proposed, we see that
for each measurement-related problem, up to $5$ different, ad-hoc
solutions existed. Implementing them in every switch is
impractical. Many of these mechanisms overlap in intent and
functionality and cannot be easily repurposed to address different
questions. To expose operators to the unnecessary complexity of
figuring out the nuances of the different variations of roughly
equivalent mechanisms makes production of new measurement code harder.
To avoid these pitfalls, we have identified a core set of primitives
that allow us to express the vast majority of common measurement tasks
and some new ones hitherto unaddressed. These primitives are
orthogonal and can be combined to express complex measurement tasks.
%Our measurement model is based on the specification of one or multiple \textit{compositions} of primitives expressing a measurement algorithm. A \textit{composition} consists of a group of primitives linked together using one of two possible operators: \seql~(\seqlsym) and \parl~(\parlsym). We will discuss their behavior in Section~\ref{sec:composition}.
They are: \match, \tagg, \ts, \sample, \counter, \bloom, \sketch, and \mmentint. We argue that switches should provide support to perform these
primitive measurement operations. As shown in \S\ref{sec:evaluation},
this set supports a wide range of measurement needs and can be mapped to hardware with a modest use of resources.

\section{\ourlang}
\label{sec:api}

A measurement task in \ourlang is expressed by combining primitives
through the sequential and parallel composition operators.
Semantically, a task is a function that processes a stream of packets
using our primitive-oriented operators (or simply primitives).
Operators take a packet as input, optionally modify state and
produce either a packet as output or the null value to stop its
processing. A task also includes flow key definitions to group
packets into flows (\eg the IP 5-tuple) and a declaration of state
variables to be used by stateful operators.

We note that measurement tasks neither interact nor influence
forwarding logic. Once a packet is consumed by a measurement task,
processing for that packet logically ends and, subsequently,  the forwarding logic
is applied until the packet leaves the switch (or gets dropped).

The input stream denoted as {\sf pkts} represents the stream of all received data packets.  The input stream denoted as {\sf ctrl} captures instead the packets received from the SDN controller. Additionally, named logical streams can be created by sampling packets; a sampled packet is duplicated and injected in a logical stream to be processed further. Each primitive executes well-defined operations on the packets in the stream as per its semantics. Our examples given below illustrate the use of our primitives on the different streams.

When describing a measurement task using MAFIA 
the user is oblivious to where the primitives are executed.
%\RM{Our model suits the architecture of programmable forwarding elements where packets are processed in a pipeline of match-action stages.}
In this paper we focus on compiling the primitives such that they 
can be executed in a programmable switch ASIC. However, our
work can be extended to compile MAFIA to other targets, such as software switches or smart NICs.

We now give an overview of \ourlang primitives (\S\ref{sec:primitives}). We then describe how primitives compose (\S\ref{sec:composition}) and discuss how we implement measurement intervals (\S\ref{sec:window}). Finally, we use the example of detecting heavy hitters to detail how \ourlang works and the nature of code that network operators would write while using our API (\S\ref{sec:byexample}).

\subsection{\ourlang Primitives}
\label{sec:primitives}

We categorize primitives in 4 classes: i) to perform filtering (\match);
ii) to manipulate packets (\tagg and \sample);  iii) to manipulate state (\ts, \counter, \sketch and \bloom); and iv) to control the measurement interval. Table~\ref{tab:table-2} presents a summary of these primitives, 
their API, and hints of their implementation in P4.
We now describe each primitive's functionality and then discuss the required resources.

\vspace{0.1em}
\smartparagraph{\match:} Filters and selects classes of packets by
parsing and inspecting the content of packet headers. Provides
conditional tests on state, allowing detection when some condition
holds.

\smartparagraph{\tagg:} Modifies or adds a header field to the packet.
Tagging is useful for piggybacking measurement data to other entities
in the network, notifying a controller, or disseminating information to
other devices.

\smartparagraph{\sample:} Makes a (logical) copy of the current
packet, separating the stream of samples from the original input
stream. Permits to send sampled packets to external entities like a controller
or a collector\cite{opensample,netsight}.

\smartparagraph{\ts:} Reads the local clock at the switch. 
The ability to derive time-related information is essential to detect
timeouts or estimate latency and packet inter-arrival times\cite{DPT}. 
Switches are not assumed to have synchronized clocks.

\smartparagraph{\counter:} Keeps track of measurable quantities such
as number of bytes, packets, etc. Counters are the standard support
for statistics in OpenFlow\cite{OpenFlow}, first realization of
SDN, and in traditional telemetry systems (NetFlow). Numerous approaches successfully leveraged
counters~\cite{HHH,openTM,OpenNetmon}.

\smartparagraph{\bloom:} Allows for efficient implementations of
membership sets. Permits to dynamically filter specific flows. Extensions to
the counting Bloom filer algorithm can also be used to store
measurement state\cite{FlowRadar,LossRadar}.

%%%%%%%%%%%%%%%%%%%%%%%%%%%%%%%%%%%%%%%%%%%%%%%%%%%%%%%%%%%%%%%%%%%%%%%%%%%%%%%%%%%%%%%%%%%%%%%%%%%%%%%%%%%%%%%%%%%% 

											% TABLE 2 %
                         
\begin{table}[t!]
{\footnotesize
%\begin{adjustwidth}{-0.15cm}{}
\begin{tabular}{| l m{11em} | m{2.25em} m{2em} p{2.25em}|}

\hline
	Primitive & 
    API & 
    \multicolumn{3}{|c|}{P4 Implementation}\\

	& 
     & 
    Tables & Actions &
    LoC

\\\hline

    \textbf{{\sf Match}} & 
    {\sf match(conditional)} & 
    1 & builtin & 
    9     

\\\hline

    \textbf{{\sf Tag}} & 
    {\sf tag(header\_field, expr)} & 
    1 & 1 & 
    9

\\\hline

    \textbf{{\sf Sample}} & 
    {\sf duplicate(stream)}  &
    %\textsl{Collect(\myquote{endpoint\_spec})} & 
    1& 1  & 22 
\\
    & 
    {\sf collect(endpoint)}  &
    
     1 &  S & S
\\\hline

    \textbf{{\sf Timestamp}} & 
    {\sf timestamp(t)} & 
    1 & 2 &
    10 
    
\\\hline

    \textbf{{\sf Counter}} & 
    {\sf set, reset} & 
    1 & 4 & 
    12 

\\\hline

    \textbf{{\sf BloomFilter}} & 
    \textit{membership}: \newline {\sf \{insert, test, reset, init\}}. &
    1 & $O(H)$ &
    $O(H)$
    \\\cline{2-2}
    &
    \textit{counting}: \newline {\sf \{set, reset, init, all, any, sum, avg, min, max\}} &
    1 & $O(H)$ &
    $O(H)$

\\\hline

    \textbf{{\sf Sketch}} & 
    \textit{pcsa/hll}: \newline {\sf \{update, test, reset\}} &
    $O(H)$ & $O(H)$ &
    $O(H)$
    \\\cline{2-2}
    &
     \textit{count-min}: \newline {\sf \{set, reset, sum, avg, min, max\}} &
    1 & $O(H)$ &
    $O(H)$
    \\\cline{2-2}
    &
     \textit{store}: \newline {\sf \{set, reset, all, any, sum, avg, min, max\}}  & 
    1 & $O(H)$ &
    $O(H)$

\\\hline

    \textbf{{\sf Window}} & {\sf window} &
    \multicolumn{3}{c|}{(variable)}

\\\hline
\end{tabular}
%\end{adjustwidth}
}
\vspace{5pt}
\caption{Measurement primitives \& API.}
\label{tab:table-2}
\end{table} 
											% TABLE 2 %

%%%%%%%%%%%%%%%%%%%%%%%%%%%%%%%%%%%%%%%%%%%%%%%%%%%%%%%%%%%%%%%%%%%%%%%%%%%%%%%%%%%%%%%%%%%%%%%%%%%%%%%%%%%%%%%%%%%% 

\smartparagraph{\sketch:} Compact data structures to hold summaries of
large datasets with provable accuracy bounds. Sketch families include
counting algorithms (count-min sketch) and cardinality estimators
(PCSA, Hyperloglog)\cite{OpenSketch,SCREAM}.

\smartparagraph{\mmentint:} Allows to specify the duration of the measurement interval (see \S\ref{sec:window}).

\vspace{0.1em}
Table~\ref{tab:table-2}'s right columns depict the number of P4 tables and actions needed to implement each primitive, with the total number of LoC (Lines of Code). The \textsl{collect} operation of the \sample primitive requires a number of actions $S$ dependent on the method of samples collection (e.g., forwarding to a monitoring server that is directly attached vs. via IP encapsulation). \bloom and \sketch require an amount of code proportional to the $H$ hash functions used; for these cases, a min-max range is given. \match does not require any custom action. \mmentint yields code dependent on the size of the structures that need to be reset (see \S\ref{sec:window}).

\subsection{Combining Primitives}
\label{sec:composition}

Primitives can be composed to express complex measurement tasks. 
We consider two forms of composition: sequential and parallel
(somewhat similar to NetKAT\cite{NetKAT}).

\paragraph{\textbf{Sequential composition}} Primitives can be composed
in serial order using the operator $\gg$.  The composition: $A \gg B
\gg \ldots \gg Z $ indicates that $A$ must be executed first, then
$B$, etc. The execution of each primitive takes into account the
effects of the previous primitive. That is, operators' side effects
(e.g., updating a counter) are made visible as soon as they execute.
This is different from NetKAT, which models policies as pure
functions. Note that operators that follow a \match are only executed
if the conditional evaluates to true.

\paragraph{\textbf{Parallel composition}} Primitives can be
parallelized via the operator $+$; the expression $A + B + \ldots + Z$
%\begin{lstlisting}[language=mafia]
%       ^$primitive_1$^ ^\textbf{+}^ ^$primitive_2$^ ^\textbf{+}^ ^$\ldots$^ ^\textbf{+}^ ^$primitive_N$^
%\end{lstlisting}
executes the primitives independently and applies multiple disjoint
measurement operations to the packet. Note that the API does not
prevent two parallelized primitives to execute concurrent operations
on the same set of state variables. In general, to prevent any
inconsistency at runtime, our intention is for read-write and
write-write conflicts to be detected ahead of time through static
analysis (left for future work).

\subsection{Measurement Interval}
\label{sec:window}

\ourlang allows the user to assign a measurement interval to any measurement task. The purpose of specifying an interval is to avoid state overflowing %\cite{FlexiblePacketProcessingPower} 
and thus corruption of measurement results. At the end of each interval, data structures are reset to their initial values before a new interval is initiated. The measurement interval is specified in time units (in the current version, in seconds) using the \textsf{window} operator.

When a measurement interval is specified, a composition has two modes of
 operation. The \textit{measuring mode}, where the primitives are invoked as 
 specified by the measurement task, and a \textit{resetting mode}, during which 
 the state of the primitive's state is reset. Given that there is a limited 
 amount of instructions that can be executed at line rate, full reset of the data 
 structures cannot be done atomically (i.e., it can take too long to reset all the data structures at once). To circumvent this limitation, all our data 
 structures support \textit{incremental reset}. When operating in resetting 
 mode, at each packet processing, incremental reset is invoked, which clears 
 a portion of the data structure. The resetting mode persists, advancing 
 incrementally each time a packet is received, until all portions of the data 
 structures have been reset; at that time a new measurement interval is initiated.
 The time required to perform the reset can be minimized by exploiting  spare resources in the switch pipeline to maximize the amount of cleared state at each packet. The actual length of the resetting phase depends on the amount of memory used by the measurement and on the packet inter-arrival times.

\begin{figure}[t]
\begin{adjustwidth}{0.5cm}{}
\begin{lstlisting}[language=mafia,caption=Two-phase heavy hitter detection with MAFIA.,captionpos=b,label={lst:mafia-hh},frame=single,numbers=left,framexleftmargin=2.25em]
flowid = Key(ip.src,ip.dest,tcp.src,tcp.dest,ip.proto)
total = Counter(width=32)
nbytes = 
 Sketch(alg="count-min",nhash=4,key=flowid,size=256,width=32)
hh = 
 BloomFilter(alg="membership",key=flowid,nhash=4,size=64)
hh_bytes = 
 HashMap(key=flowid,size=1024,type=Counter(width=32))
window(mment_interval)
// Heavy hitter detection.
pkts
  ^$\seqlsym$^ match(pkt.input_port == ^\textsl{PORT}^) 
  ^$\seqlsym$^ total.set(total + pkt.size)
  ^$\seqlsym$^ (( match(!hh.test()) 
           ^$\seqlsym$^ nbytes.set(nbytes + pkt.size)
           ^$\seqlsym$^ match(nbytes.min() / total > ^$\gamma$^) 
           ^$\seqlsym$^ hh.insert()
           ^$\seqlsym$^ hh_bytes.set(nbytes.min())
           ^$\seqlsym$^ duplicate(hh_alarms) )
       ^$\parlsym$^
        ( match(hh.test()) ^$\seqlsym$^ hh_bytes.set(hh_bytes + pkt.size)))
// Alarms sent to the SDN controller.
hh_alarms
  ^$\seqlsym$^ tag(ipv4.checksum, nbytes.min()) ^$\seqlsym$^ collect(^\textsl{CONTROLLER}^)
// Control traffic to retrieve heavy hitters volume.
ctrl
  ^$\seqlsym$^ match(pkt.request==^\textsl{HH\_VOLUME}^) ^$\seqlsym$^ duplicate(get_hh_volume)
get_hh_volume
  ^$\seqlsym$^ tag(pkt.hh_volume, hh_bytes) ^$\seqlsym$^ collect(^\textsl{CONTROLLER}^)
\end{lstlisting}
\end{adjustwidth}
\end{figure}

\subsection{MAFIA by Example}
\label{sec:byexample}

We present the flavor of the \ourlang API through a use case and
discuss the abstractions on which it relies, as well as detailing the
behavior of the measurement primitives.

Consider the problem of identifying heavy hitters, \ie flows that
consume more than a fraction $\gamma$ of link capacity. Typical
approaches to this problem consist of installing forwarding rules
associated with counters to monitor flows~\cite{HHH,DREAM}. However,
limited switch memory makes it impossible to install a separate rule
for each flow; the common strategy is to monitor a set of aggregates
(\ie grouping classes of flows at coarse granularities) and then
zooming in on the ones most likely to contain heavy flows. %(\ie splitting the rules into multiple, more specific ones).
Unfortunately,
this technique introduces a detection delay since it counts in several
consecutive time intervals. Other approaches adopt approximate algorithms such as the count-min sketch to
approximately count the size of flows~\cite{OpenSketch,SCREAM}. A
monitoring server can then retrieve the sketch data from the switch to
compute the identifiers of the top-k largest flows. Although sketches
can provide provable bounds on the estimation error, collisions between different flows are
workload-dependent and hard to predict.

Instead, we combine these approaches to obtain fast detection
with accurate results. \ourlang allows to do so by flexibly
composing a few primitive operations.
%Listing~\ref{lst:mafia-hh} presents the code addressing this heavy hitter measurement.
The algorithm works in two phases. First, it identifies potentially
large flows using a count-min sketch. After a flow has been detected
as a possible heavy hitter (by checking against a user-defined
threshold), its identifier is encoded in a Bloom filter. The filter
keeps track of the set of suspected flows, whose packets will then be
monitored using exact counters instead of using a sketch. Next, we
send alarms to a controller whenever suspected flows are inserted in
the filter.  The corresponding \ourlang code
(Listing~\ref{lst:mafia-hh}) consists of two main parts: i)
measurement state declarations and ii) the composition of primitives.

\subsubsection{\textbf{Measurement state}}
Lines 1-8 present an example of state declaration for the considered
measurement.

\paragraph{Keys}
Keys allow us to group packets to flows and map them to measurement
state. Each state variable can be configured to maintain multiple
instances of the same primitive's state. This is done using a hash
map that, through a hash function of the key, obtains an index at which
to hold flow state. Hash maps are meant to be simple container
abstractions and do not handle potential hash collisions on different
keys in our implementation. Eviction techniques for collision handling such as the one adopted in
\cite{Marple} are amenable to be used within our approach. Bloom filters and
sketches adopt a set of hash functions to implement the mapping of packets to state.
% Multiple flows can be monitored using the State primitives can
% declare multiple \ourlang relies on this notion of keys to identify
% and group flows logically.

\paragraph{State}
State is important in a wide range of measurement needs, including
tracking traffic conditions, filtering events and maintaining
statistics or counters. State is maintained in the data plane and
updated at line rate.

State is declared as named variables for stateful primitives. In the
example (Lines 2-8), we have i) a counter {\sf total}, which is used
to track the total volume of flows received on an input port; ii) a
sketch {\sf nbytes} using a count-min algorithm to approximately
measure the volume of a set of flows; iii) a membership Bloom
filter {\sf hh} holding the set of flows to be monitored by exact
counters; and iv) an hash map of counters {\sf hh\_bytes} used to
track the volume of heavy hitters.

Declaration of state variables is required by primitives manipulating
them (\ie \ts, \counter, \sketch and \bloom). Variables of these types
all require some parameters, such as the {\sf width} of counters (in
bits), the number of hash functions {\sf nhash} for sketches and Bloom
filters, etc.

\ourlang does not mandate how state is consumed. In certain contexts,
the maintained state serves as a filter to pick up interesting packets
or flows that need to be collected at monitoring servers. In other
contexts, the maintained state (\eg flow counters, volume sketches) is
the information of interest; this information can be queried in-band
(as in \cite{TPP}) using \tagg, or via switch-specific APIs
(beyond our scope).

\subsubsection{\textbf{Composition of primitives}}

A composition of primitives is the core part of a measurement task in
\ourlang. In Listing~\ref{lst:mafia-hh}, two compositions of
primitives (Lines 12-21 and 23-24) implement the heavy hitter
monitoring and controller alarms, respectively. The entry point is at
Line 11.

First, we select (Line 12) from the {\sf pkts} stream all packets from
a given input port ({\sf PORT}) using the \match primitive with a
conditional on the {\sf pkt}'s input port field.  Recall packets are
parsed into tuples and we allow expressions on packet header fields
and metadata via intuitive keywords (\eg {\sf ipv4.src} for the IP
source address, {\sf pkt.size} for the packet size). The available packet headers are the ones that derive from the switch packet parsing procedure. Then, we
maintain (Line 15) a {\sf total} count of traffic volume ingressed.
%at {\sf PORT}.

Next, depending on whether the current packet belongs (Line 14) to the
set of heavy hitters ({\sf hh}) or not (Line 21), one of two things
happen. As these are independent, we express them as a parallel
composition guarded by the (mutually exclusive) tests on the {\sf hh}
\bloom.  Lines 15-19 measure flows not (yet) suspected to be heavy
hitters, while Line 21 measures the heavy hitters.  In the first
case (non heavy hitter), we update (Line 15) the flow's volume in the
{\sf nbytes} \sketch and then we query (Line 16) the count-min
sketch to check whether the flow's estimated bandwidth utilization is
above $\gamma$. We insert (Line 17) flows that exceed the threshold
into the \textsl{hh} Bloom filter and we raise (Line 19) an alarm for
each such flow, by duplicating the packet into the {\sf hh\_alarms}
stream.  In the second case (heavy hitter), we update (Line 21) the
flow's volume tracked by an exact counter in {\sf hh\_bytes},
which is initialized in Line 19 with the current value of the
sketch when the heavy hitter flow is flagged.

The last part of the code, on Lines 23-24, handles the duplicated
packets generated whenever a flow is detected as heavy hitter. We tag
the packets with the current flow volume from the sketch and forward them to the controller node processing these alarms.
The counter values of flows identified as heavy hitters can be queried by the controller using control requests: Lines 26-29 handle this case.  We match control traffic for specific requests (Line 34) and, as a response, we tag the queried value (Line 29). Note that the controller needs to know the counters where the queried value is. We use a straightforward solution: given that flagged flows are sent to the controller as alarms, the controller can use packet crafting techniques\cite{ATPG} to forge the necessary header values when querying the switch state.

This example, addressed with only 29 lines of code, is able to improve
measurement efficacy when compared to the common black-box solutions using only
sketch-based algorithms or exact counters. It merges the benefit of
small memory requirements of sketches to identify which flows should
be monitored (reducing the number of expensive counters to be
employed) with the precision of exact counters to determine the volume
of heavy flows (thus eliminating estimation errors due to hash
collision in the sketch). {Collisions in the bloom filter maintaining the current set of heavy hitters that may pollute the exact counters of the heavy flows can be rendered
unlikely at a small cost to memory.}

\section{Implementation}
\label{sec:implementation}

We implement \ourlang as a small domain specific
language embedded in Python. Although our approach is
not tied to any particular programmable forwarding element
implementation, as backend we target PISA (Protocol-Independent Switch
Architecture)~\cite{ForwardingMetamorphosis} switches programmed in P4. Our implementation consists of a compiler (around 4,500 lines of code in Python) that takes a MAFIA measurement as input and produces an equivalent P4 program.

\begin{figure}
  \includegraphics[width=\linewidth]{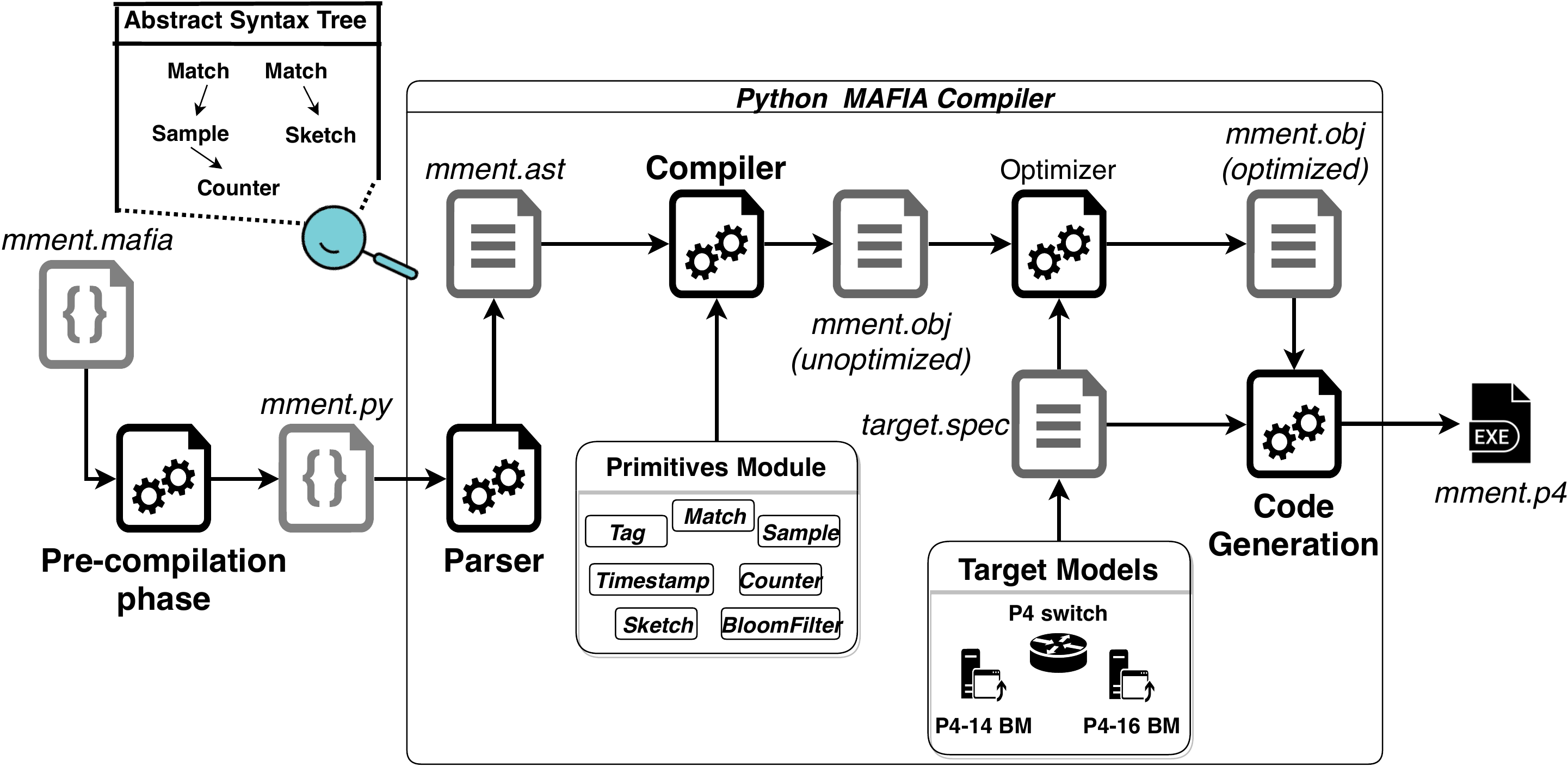}
  \caption{Overview of compilation process.}
  \label{fig:mafia-architecture}
\end{figure}

The compilation process is illustrated in
Figure~\ref{fig:mafia-architecture}. We follow a 
compilation sequence composed of the five phases enumerated below. For brevity, we
highlight the salient details.  A pre-compilation phase translates a
\ourlang code into Python. Then, the corresponding AST (Abstract
Syntax Tree), where operators are nodes of the tree, is built.  The
compiler then produces an intermediate representation of unoptimized
P4 code by analyzing the primitive composition and translating it into
a set of tables and actions. These are the P4 processing blocks (packet functions) of \ourlang code, whose 
execution is controlled via a sequence of P4 table calls that is determined by
the compiler analyzing how primitives were composed via the \parlsym~and \seqlsym~operators.
Then, an optimization phase takes as
input a target's model specification, and produces the code optimized for the
target's architecture. This optimization phase accounts for the different target capabilities and handles differences in syntax of dialects of P4 (P4-14, P4-16).
Finally, the process generates P4 code for tables, actions and registers. Extending our solution to target different architectures, software switches or smart NICs, is left as future work. Table~\ref{tab:table-2} shows the number of P4-14 lines of code  and the number of tables and actions necessary to implement each primitive.

\section{Evaluation}
\label{sec:evaluation}

We evaluate our approach along three axes. We assess the
expressiveness of \ourlang by implementing 13 measurement tasks
that are routinely done along with some new ones. We then compare the
efficiency of the P4 code generated by our compiler by contrasting the
complexity of \ourlang implementation of these tasks with a manual
implementation. Finally, we show the feasibility of our approach by
assessing the hardware resource required to deploy each considered
measurement use case on programmable switches.

\subsection{Expressiveness}
\label{sec:expressiveness}

The measurement tasks are enumerated in Table~\ref{tab:table-3}.
The \ourlang code for each use case can be found in Appendix~\ref{sec:appendix}.
%The corresponding domain-specific code, for all these measurements is presented in Appendix~\ref{sec:appendix}. 
% Due to space constraints we do not show the corresponding code for
% all (presented in a technical report companion to this paper~\cite{mafia-tech-report}, 
Here, we focus on three novel measurements that demonstrate the re-usability and
composability of our primitives: i) identification of congested flows
by improving a state-of-the-art solution offered by INT (In-band
Network Telemetry)~\cite{INT}; ii) measurement of path changes and iii)
and monitoring of the distributed coordination of a novel path update
protocol, ez-Segway~\cite{ez-segway}. Some of these tasks require deploying
measurement tasks in different switches whose identity is assumed to
be known at deployment time. As mentioned, we show capability of \ourlang
in some new measurement tasks beyond well-known ones.

%An opportunity for future work would consist in primitives placement planning for a network-wide setting, and consider dynamically deployed measurements.

\paragraph{\textbf{Top-k Congested Flows}}

Detecting flows that are experiencing large queuing delays in the
network is important to guarantee quality of service and meet service
level agreements. INT collects hop-by-hop information (\eg
queue occupancy, hop latency) for each packet by using a custom
header, inserted to the packet at the network ingress point and
stripped at egress point. However, potentially large per-hop
information needs to be exported in order to be used for analysis.
%INT's ability to collect, potentially large, per-hop information is traded off against the use of a non-standard header which may or may not be handled correctly by devices. 
We can improve on this by using a simple composition of
primitives and provide a stateful algorithm identifying
the most congested flows.
%The code requires only to reserve a single header field of the packet (\eg the IP identification field), as opposed to INT which requires a non-stander header. 
Listing~\ref{lst:mafia-topkc} shows the \ourlang code for this
measurement task. It has three code segments to be installed at
different switches. 1) The first-hop switch marks the packet's IP
ToS field to indicate \emph{whether} the measurement should be applied,
and tags the IP ID field with the current queue occupancy level. 2)
Intermediate switches check if the packet has been marked and update
the tag by summing the local queue occupancy. 3) The last switch
records the total amount of queuing encountered along the path
using a count-min \sketch, which can be periodically queried to compute the top-k elements. Another \sketch using the same flow key instead tracks the total
number of changes happening over time. We could add a
\counter tagged in packets to track the number of hops traversed by each packet, useful to compute an
average value of queue occupancy at each hop. A similar measurement
could also sum up the queuing time experienced at each hop, using
\ts.

% \begin{figure}[t]
% \begin{adjustwidth}{0.5cm}{}
% \begin{lstlisting}[language=mafia,caption=Identifying the top-k congested flows,captionpos=b,label={lst:mafia-topkc},frame=single,numbers=left,framexleftmargin=2.25em]
% window(mment_interval)
% // Code executed at first hop:
% pkts 
%   ^$\seqlsym$^ tag(ipv4.tos, ipv4.tos | 0x1) 
%   ^$\seqlsym$^ tag(ipv4.id, pkt.in_queue_length))
% 
% // Code executed at intermediate hops:
% q_len = Counter(width=32);
% pkts
%   ^$\seqlsym$^ match(ipv4.tos & 0x1 == 0x1) 
%   ^$\seqlsym$^ q_len.set(ipv4.id + pkt.in_queue_length)
%   ^$\seqlsym$^ tag(ipv4.id, q_len)
% 
% // Code executed at last hop:
% flowid = Key(ip.src,ip.dest,tcp.src,tcp.dest,ip.proto)
% total_pkts = 
%   Sketch(alg="count-min",key=flowid,nhash=4,size=1024,w=32)
% path_q_len = 
%   Sketch(alg="count-min",key=flowid,nhash=4,size=1024,w=32)
% 
% pkts
%   ^$\seqlsym$^ match(ipv4.tos & 0x1 == 0x1)
%   ^$\seqlsym$^ total_pkts.set(total_pkts + 1)
%  ^$\seqlsym$^ path_q_len.set(path_q_len + ipv4.id)    
% \end{lstlisting}
% \end{adjustwidth}
% \end{figure}
% 

\begin{figure}[t]
\begin{adjustwidth}{0.5cm}{}
\begin{lstlisting}[language=mafia,caption=Identifying the top-k congested. flows,captionpos=b,label={lst:mafia-topkc},frame=single,numbers=left,framexleftmargin=2.25em]
window(mment_interval)
// Code executed at first hop:
pkts ^$\seqlsym$^ tag(ipv4.tos, ipv4.tos | 0x1) 
     ^$\seqlsym$^ tag(ipv4.id, pkt.in_queue_length))
// Code executed at intermediate hops:
q_len = Counter(width=32);
pkts ^$\seqlsym$^ match(ipv4.tos & 0x1 == 0x1) 
     ^$\seqlsym$^ q_len.set(ipv4.id + pkt.in_queue_length) 
     ^$\seqlsym$^ tag(ipv4.id, q_len)
// Code executed at last hop:
flowid = Key(ip.src,ip.dest,tcp.src,tcp.dest,ip.proto)
total_pkts = 
  Sketch(alg="count-min",key=flowid,nhash=4,size=1024,w=32)
path_q_len = 
  Sketch(alg="count-min",key=flowid,nhash=4,size=1024,w=32)
pkts ^$\seqlsym$^ match(ipv4.tos & 0x1 == 0x1)
     ^$\seqlsym$^ total_pkts.set(total_pkts + 1)
     ^$\seqlsym$^ path_q_len.set(path_q_len + ipv4.id)    
\end{lstlisting}

\begin{lstlisting}[language=mafia,caption=Measuring flow path changes.,captionpos=b,label={lst:mafia-pathchange},frame=single,numbers=left,framexleftmargin=2.25em]
// Code to be executed at intermediate switches
location = Key(pkt.input_port, switch.id, pkt.output_port)
location_bf = 
      BloomFilter(alg="membership",key=location,nhash=4,size=32)
pkts ^$\seqlsym$^ location_bf.init(ipv4.checksum)
     ^$\seqlsym$^ location_bf.set()
     ^$\seqlsym$^ tag(ipv4.checksum, location_bf)
     ^$\seqlsym$^ location_bf.reset()
// Code to be executed at the packet's last hop
flowid = Key(ip.src,ip.dest,tcp.src,tcp.dest,ip.proto)
paths_sketch = 
      Sketch(alg="store",key=flowid,nhash=4,size=256,width=32)
n_change_sketch = 
      Sketch(alg="countmin",key=flowid,nhash=4,key=flowid,size=256)
window(mment_interval)
pkts ^$\seqlsym$^  match(!paths_sketch.any(ipv4.checksum))
     ^$\seqlsym$^  paths_sketch.set(ipv4.checksum)
     ^$\seqlsym$^  n_change_sketch.set(n_change_sketch + 1)
\end{lstlisting}

\begin{lstlisting}[language=mafia,caption=Monitoring the ez-Segway~\cite{ez-segway} protocol.,captionpos=b,label={lst:mafia-pathchangetime},frame=single,numbers=left,framexleftmargin=2.25em]
// Code to be executed on all switches updating rules
change_ts = Timestamp();
l_clock = Counter(width=8);
pkts ^$\seqlsym$^ match(segway_header.msg == GoodToMove) 
     ^$\seqlsym$^ l_clock.set(max(l_clock + 1, segway_header.ts))
     ^$\seqlsym$^ tag(segway_header.ts, l_clock)
     ^$\seqlsym$^ duplicate(end_of_update) 
end_of_update ^$\seqlsym$^ timestamp(change_ts) 
     ^$\seqlsym$^ tag(segway_header.time, change_ts)
     ^$\seqlsym$^ tag(segway_header.ts, l_clock)
     ^$\seqlsym$^ collect(^\textsl{SEGWAY\_CONTROLLER}^)
\end{lstlisting}
\end{adjustwidth}
\end{figure}

\paragraph{\textbf{Path Changes}}

Recently proposed load balancing mechanisms, such as flowlet
switching~\cite{HULA}, autonomously cause
path changes without any coordination with a controller. One may want to monitor such mechanisms to understand how often a
given flow changes its path. We show how such a task can be realized
using \ourlang in Listing \ref{lst:mafia-pathchange}.
%SDN allows for faster and frequent changes in network behavior. As each flow could be routed through many available paths, the controller may change the forwarding policies multiple times to avoid bottlenecks and optimize resources. This use case focuses on answering how often a given flow changes its path which will help with both traffic steering decisions and performance debugging. Note that although the controller might know about deployed forwarding changes, it may not have sufficient resources for carrying out measurements. Moreover, updates may be delayed due to network delays in reaching the switch or to meet dependencies\cite{ez-segway}. Forwarding rules may be time-based, expiring after timeouts, thus causing inconsistencies with the control plane view. Load balancing mechanisms such as flowlet switching~\cite{FlexiblePacketProcessingPower} as well cause path changes without coordination with the SDN controller.
To encode the path followed by a packet, we use a \bloom
to store the packet location (\ie the current switch ID
and port), which is tagged into the packet's IP ID
field. The filter is updated at every hop, resulting
in a compact representation of the path. At the last hop, we collect
the tag, and save it inside a \sketch, which maintains
the identifiers of the packet's path.

%\pl{we collect the tag, and save it inside a sketch data structure, which thus maintains the identifiers of the packet's path for a set of flows. } \RM{we use a per-flow Sketch to maintain the last path followed: we use a general implementation of a sketch algorithm to store a path's Bloom filter tag}\mcnote{Not sure what this means}. 
The measurement checks, at every packet, if the carried path tag value
is found in any entry held in the sketch (\ie holding the flow status
at the previous packet). If not, a path change is detected, and a
count-min sketch tracking the amount of changes is updated.  The
controller can fetch the data at desired frequency to learn about path
changes.

\paragraph{\textbf{Path Change Coordination and Latency}} 

Implementing path changes in SDN networks often involves updating
forwarding rules in multiple switches. Lack of proper coordination in
these changes may result in transient inconsistencies such as black
holes, loops, or link overloads. There has been significant research
effort in techniques to provide consistent forwarding updates in
SDN. We focus our attention on ez-Segway\cite{ez-segway}, a technique
that shortens the time required to perform consistent path changes by
implementing a coordination mechanisms among the switches that requires the
exchange of ``GoodToMove'' messages in a given order. We now describe a
measurement that captures the partial order by which the
``GoodToMove'' messages are exchanged during the reconfiguration and
also the time at which these messages are received by each switch
involved in the task. This measurement can be used to find bugs in the
coordination algorithm that may prevent the path change protocol from
terminating and also to assess how long it takes to execute.  The
\ourlang code is shown in Listing~\ref{lst:mafia-pathchangetime}. We
detect the receipt of a ``GoodToMove'' message and generate a duplicate to be sent back to the
controller. Both a local \ts and a logical
clock are recorded and tagged before the copy is sent back to the
controller via sampling. The use of logical clocks allows the
controller to build a causal graph of the deployed updates, providing
an execution log that can be queried for debugging and
verification. If the controller's and switches' clocks are
synchronized (\eg via NTP), real timestamps permit the controller to
estimate the time between change deployments and their actual
occurrence.

%Note that the detection of the \myquote{GoodToMove} packet may, or may not, be feasible to be done in the fast path of the switch (\eg control packet being encrypted). However, we do not make any specific assumption on where our primitives approach may operate. While the fast path is best for measurements at line rate without  disrupting switch operations, we do not preclude switch CPUs from executing our primitives. In future work we will explore the benefits of executing in the slow path. As the example above shows, it may be useful to have a way to measure the control traffic dedicated to the coordination of the data plane from the control plane.

%\begin{figure}[t]
%\begin{adjustwidth}{0.5cm}{}
%\begin{lstlisting}[language=mafia,caption=Measuring the latency to path steering. \newline The measurement relies on the ez-segway~\cite{ez-segway} path update protocol,captionpos=b,label={lst:mafia-pathchangetime},frame=single,numbers=left,framexleftmargin=1.775em]
%// Code to be executed at the packet's last hop
%change_ts = Timestamp();
%pkts
%  ^$\seqlsym$^ match(segway_header.msg == GoodToMove) 
%  ^$\seqlsym$^ duplicate(end_of_update) 
% end_of_update
 % ^$\seqlsym$^ timestamp(change_ts)
 % ^$\seqlsym$^ tag(segway_header.time, change_ts) 
%\end{lstlisting}
%\end{adjustwidth}
%\end{figure}
%\input{api-code/mafia-path-latency}

%%%%%%%%%%%%%%%%%%%%%%%%%%%%%%%%%%%%%%%%%%%%%%%%%%%%%%%%%%%%%%%%%%%%%%%%%%%%%%%%%%%%%%%%%%%%%%%%%%%%%%%%%%%%%%%%%%%% 

											% TABLE 3 %
                         
\begin{table}[t!]
\begin{center}
%\begin{adjustwidth}{-0.5cm}{}
%\begin{adjustwidth}{-0.85cm}{}
{\footnotesize
%\begin{tabular}{|p{15em}p{15em} p{4.5em} p{3.7em} p{4.5em}|}
\begin{tabular}{|p{5em}p{11.25em} p{3.30em} p{2.25em} p{2.25em}|}
\hline
  &  
  & 
  \multicolumn{3}{c|}{P4 LoC} 
\\
   Measurement &  
  &
   (Manual) &
   \multicolumn{2}{c|}{(Compiler)}
\\
  Use case &  
  API: Primitives & 
  &
  raw & opt.
\\\hline\hline

  \textbf{\textit{Flow volume and \newline duration}} &  
  3 $\times$ Match; \newline 
  3 $\times$ Counter HashMap; \newline 
  2 $\times$ Timestamp HashMap; & 
  %8 &
  121 &
  %185 / 
  185 & 146 \newline {\tiny ($+20\%$)} %&
  %20
\\\hline

  \textbf{\textit{Approximate \newline flow volume}} & 
  1 $\times$ Match; \newline 
  1 $\times$ Sketch (count-min) & 
  %2 &
  107 &
  120 & 120 \newline {\tiny ($+12\%$)} %
  %&
  %10
\\\hline

  \textbf{\textit{Flow \newline cardinality}} & 
  1 $\times$ Match; \newline 
  1 $\times$ Sketch (PCSA) & 
  %2 &
  86 &
  92 & 92 \newline {\tiny ($+6\%$)} %%& 
  
  %11
\\\hline

  \textbf{\textit{Flow \newline cardinality}} & 
   1 $\times$ Match; \newline 
   1 $\times$ Sketch (HyperLogLog) & 
  %2 &
  96 &
  102 & 102 \newline {\tiny ($+2\%$)} %%&
  %9
\\\hline

  \textbf{\textit{Counter \newline thresholds}} & 
  5 $\times$ Match; \newline 
  2 $\times$ Counter HashMap; \newline 
  2 $\times$ Sample & 
  %11 &
  139 &
  %193 / 
  193 & 170 \newline {\tiny ($+22\%$)} %%&
  %12
\\\hline

  \textbf{\textit{Stochastic \newline sampling}} & 
  2 $\times$ Match; \newline 
  1 $\times$ Tag; \newline 
  1 $\times$ Sample; & 
  %4 &
  103 &
  %126 / 
  126 & 118 \newline {\tiny ($+14\%$)} %%&
  %10
\\\hline

  \textbf{\textit{Deterministic \newline sampling}} & 
  5 $\times$ Match; \newline
  3 $\times$ Counter HashMap; \newline 
  1 $\times$ Tag; \newline 
  1 $\times$ Sample; & 
  %13 &
  131 &
  %207 / 
  207 & 167 \newline {\tiny ($+27\%$)} %%& 
  %19
\\\hline

  \textbf{\textit{Postcard \newline generation}} & 
  2 $\times$ Match; \newline 
  4 $\times$ Tag; \newline 
  1 $\times$ Sample; & 
  %7 &
  94 &
  %121 / 
  121 & 101 \newline {\tiny ($+7\%$)} %%&
  %9
\\\hline

  \textbf{\textit{Trajectory \newline encoding}} & 
  5 $\times$ Match; \newline
  1 $\times$ BloomFilter; \newline 
  1 $\times$ Timestamp+HashMap ; \newline 
  6 $\times$ Tag;  \newline 
  1 $\times$ Sample; \newline  
  1 $\times$ Counter; & 
  %17 &
  244 &
  %299 / 
  299 & 260 \newline {\tiny ($+6\%$)} %%&
  %13
\\\hline\hline

\textbf{\textit{Two-phase \newline heavy \newline hitter}} & 4 $\times$ Match; \newline 
  1 $\times$ Counter; \newline 
  1 $\times$ Counters HashMap; \newline 
  1 $\times$ Sketch (count-min); \newline 
  1 $\times$ BloomFilter; & 
  %11 &
  261 &
  %345 / 
  345 & 281 \newline {\tiny ($+8\%$)} %%&
  %25
\\\hline

\textbf{\textit{Top-k \newline congested flows}} & 
  3 $\times$ Match; \newline 
  1 $\times$ Counter; \newline 
  2 $\times$ Sketch (count-min); \newline 
  3 $\times$ Tag; & 
  %8 &
  198 &
  %240 / 
  240 & 204 \newline {\tiny ($+3\%$)} %%&
 % 14
\\\hline

\textbf{\textit{Path changes}} & 
  3 $\times$ Match; \newline 
  1 $\times$ Sketch (count-min); \newline 
  1 $\times$ Sketch; \newline 
  1 $\times$ BloomFilter; \newline 
  1 $\times$ Tag; & 
  %9 &
  325 &
  %389 / 
  389 & 345 \newline {\tiny ($+6\%$)} %%&
 % 21
\\\hline

\textbf{\textit{Path change \newline latency}} & 
  2 $\times$ Match; \newline 
  1 $\times$ Timestamp; \newline 
  1 $\times$ Sample; \newline 
  1 $\times$ Tag; & 
  %5 &
  38 &
  44 & 41 \newline {\tiny ($+8\%$)} %%&
%  8
\\\hline

\end{tabular}
}
%  \caption{Measurement use cases and the set of primitives from our API required to implement them. Comparison between the size of an handcoded P4 implementation and the compiler-generated equivalent. \color{blue}{2.3.11 - The flow cardinality use cases has been address with two different variations of sketches\cite{SketchPCSA}\cite{SketchHyperLogLog}}. The LoC for the unoptimized (raw) and optimized compiled program are shown.}
  \caption{Use cases.}
  \label{tab:table-3}
%\end{adjustwidth}
\end{center}
\end{table}
											% TABLE 3 %

%%%%%%%%%%%%%%%%%%%%%%%%%%%%%%%%%%%%%%%%%%%%%%%%%%%%%%%%%%%%%%%%%%%%%%%%%%%%%%%%%%%%%%%%%%%%%%%%%%%%%%%%%%%%%%%%%%%% 

\subsection{Efficiency} 

Table~\ref{tab:table-3} shows the number of \ourlang  primitives required to
specify each of the considered use cases. The table also compares the number of LoC needed to implement the measurement manually in P4 and the resulting size of the compiler-generated code (optimized and not).  Both the compiler-generated code and the manual implementation instantiate the same amount of state, which means that our compiler is efficient and does not introduce state overhead. {The flow cardinality use cases has been address by two different variations of sketches\cite{SketchPCSA}\cite{SketchHyperLogLog}}. 
The code complexity of the optimized version is comparable to hand-written code: the P4 LoCs produced by \ourlang are, depending on the task, $3\%$ to $27\%$ larger than those of the code written manually in P4. The use cases using a single sketch (approx. flow volume and cardinality) are always optimized by our compiler.

%Header definitions, the parser specification and register declarations are not included in the LoC metric, since both the P4 manual implementation and the compiler's generated code would need identical code.
% DIFF
%\RM{as they are necessary independently} \mcnote{Previous two words mean nothing to me.} \pl{since both the P4 manual implementation and the compiler's generated code would need identical code. }
%Given that our compiler does not leverage any particular optimization technique, the resulting P4 program is larger than that of a manual implementation, where it is possible to optimize the number of match-action tables by grouping them together, leveraging high-level knowledge of the measurement operations.
%\mcnote{Need to revise this in light of the optimizer.}
%\mcnote{Implications of larger P4 LoCs are unclear.}

The low number of primitives employed in each use case demonstrates
that it is possible to express the measurements concisely. With a
handful of primitive invocations, it is possible to express measurement
techniques that would otherwise require significant coding effort. Our
API is able to convey the measurement intent and describe the
operational steps involved allowing network operators to focus on the
measurement needs to be carried out rather than issues arising from
using a low-level language such as P4.
%DIFF
%\RM{Due to space limit, we do not show the complete code using our API for these use cases.} \mcnote{Already said.}
Compilation times for each of the considered examples is always below 25 ms and is negligible
compared to the effort of implementing several ad hoc solutions in
P4. Our primitive-oriented approach benefits from the reusability of a
simple set of measurement functionality.
%Their full specification can be analyzed in our technical report\cite{mafia-tech-report}.

\subsection{Feasibility} 

To demonstrate the deployment feasibility of our approach, we estimate the resource requirements of a hardware switch\footnote{We are not allowed to report results for the actual resource requirements on a Barefoot's Tofino switch chip due to a confidentiality agreement.} for the measurements presented in Table~\ref{tab:table-3}, following the same methodology as in previous work~\cite{Marple}. The metrics we consider are: i) the depth of the switch processing pipeline (\ie number of match-action stages); ii) the width of the pipeline (\ie the maximum number of parallel computation that needs to be performed in a single stage); and iii) the total number of processing atoms that each measurement occupies in the switch pipeline.

We model each of the measurements using the Banzai machine model~\cite{Domino} and compile the Banzai code using its compiler, Domino. A Banzai machine comprises of stateless atoms, which are able to perform binary operations (arithmetic, logic, and relational) on pairs of packet fields, and one stateful atom, capable of accessing and updating the switch registers. 
%\RM{Banzai models different types of atoms depending on their complexity and features.} 

Table~\ref{tab:table-4} shows that all of the measurements considered can be mapped to hardware with a modest use of resources. The most complex measurement requires 9 pipeline stages and 13 concurrent operations per stage, for a total of 49 atoms. To put this in context, current programmable switch chips~\cite{ForwardingMetamorphosis}, like Barefoot's Tofino (24 stages and up to 63 actions per stage), already fulfill these requirements.

\ourlang expects a target switch to make a certain amount of memory available
for primitives. Current programmable switch architectures support 0.5 - 32 Mb 
of memory for each stage [8], organized in register arrays. Memory 
employed in \ourlang is mapped to registers by our compiler.

%\rvw{3.1.1}{\ourlang assumes a target switch makes available a certain amount of memory to be employed by primitives. Current programmable switch architectures support 0.5 –- 32 Mb of memory for each stage\cite{ForwardingMetamorphosis}, organized in register arrays. Stateful memory employed in \ourlang is mapped naturally to registers by our compiler. }

%DIFF
%\RM{Given that a PISA architecture\cite{ForwardingMetamorphosis} offers 32 stages and can provide hundreds of parallel instructions per stage, the measurement we showcased occupy a fairly modest amount of resources.} \mcnote{Is that the limit of PISA architecture or a particular instantiation?}  \pl{Current PISA\cite{ForwardingMetamorphosis} architectures already fulfill such requirements (\eg the hardware Tofino switch has 24 stages with up to 63 actions per stage).}

%%%%%%%%%%%%%%%%%%%%%%%%%%%%%%%%%%%%%%%%%%%%%%%%%%%%%%%%%%%%%%%%%%%%%%%%%%%%%%%%%%%%%%%%%%%%%%%%%%%%%%%%%%%%%%%%%%%% 

											% TABLE 3 %
                         
\begin{table}[t!]
\begin{center}
%\begin{adjustwidth}{-0.125cm}{}
%\begin{adjustwidth}{-0.1cm}{}
{\footnotesize
\begin{tabular}{|p{11em} p{2.5em} p{2.5em} p{2.5em} p{5.5em}|}
\hline
  Measurement &  
  Pipeline depth & 
  Pipeline width &
  Num. Atoms & 
  Banzai \newline Atom Type
\\\hline\hline

  \textbf{\textit{Flow volume and duration}} &  
  4 & 
  4 &
  11 &
  Sub
\\\hline

  \textbf{\textit{Approximate flow volume}} & 
  4 & 
  5 &
  18 &
  RAW
\\\hline

  \textbf{\textit{Flow cardinality}} & 
  3 & 
  3 &
  6 &
  RW
\\\hline

  \textbf{\textit{Flow cardinality}} & 
  3 & 
  2 &
  4 &
  RW
\\\hline

  \textbf{\textit{Counter thresholds}} & 
  5 & 
  2 &
  9 &
  If-Else-RAW
\\\hline

  \textbf{\textit{Stochastic sampling}} &
  3 & 
  1 &
  3 &
  If-Else-RAW
\\\hline

  \textbf{\textit{Deterministic sampling}} & 
  6 & 
  2 &
  8 &
  Pairs
\\\hline

  \textbf{\textit{Postcard generation}} & 
 1 & 
 5 &
 5 &
 RW
\\\hline

  \textbf{\textit{Trajectory encoding}} & 
 6 & 
 3 &
 8 &
 RW
\\\hline\hline

\textbf{\textit{Two-phase heavy hitter}} & 
8 & 
12 &
41 &
If-Else-RAW 
\\\hline

\textbf{\textit{Top-k congested flows}} & 
9 & 
6 &
38 &
If-Else-RAW 
\\\hline

\textbf{\textit{Path changes}} & 
9 & 
13 &
49 &
If-Else-RAW 
\\\hline

\textbf{\textit{Path change latency}} & 
 4 & 
 2 &
 5 &
 RW
\\\hline

\end{tabular}
}
  \caption{Resource requirements of use-cases.}
%  : pipeline depth, pipeline width and total number of atoms. Assessed using the Domino/Banzai\cite{Domino} machine model.}
  \label{tab:table-4}
%\end{adjustwidth}
\end{center}
\end{table}
											% TABLE 3 %

%%%%%%%%%%%%%%%%%%%%%%%%%%%%%%%%%%%%%%%%%%%%%%%%%%%%%%%%%%%%%%%%%%%%%%%%%%%%%%%%%%%%%%%%%%%%%%%%%%%%%%%%%%%%%%%%%%%% 

\section{Related Work}
\label{sec:related-work}

%\mcnote{Reads like an intro/motivation paragraph instead of rel. work}

%Our work builds upon a rich literature in network measurements, which is too vast to mention exhaustively here.
%We already referenced in Section~\ref{sec:scenarios} an extensive list of work that inspired out choice of primitives.

Our selection of primitives is inspired by many successful measurement approaches \cite{HHH, openTM, OpenNetmon, OpenSketch, UnivMon} that demonstrated the value of maintaining exact packet or byte counters or approximate estimates via sketches or Bloom filters, sampling information, exporting information via packet tagging, etc.
These ideas were applied successfully in a range of contexts from detecting heavy-hitters to estimating the traffic matrix, to measuring throughput, latency, and packet loss.

OpenFlow~\cite{OpenFlow} supports traffic monitoring by automatically associating packets and bytes counters to flow match rules installed by the controller. These have been used to perform heavy-hitter detection, traffic matrix estimation, throughput, latency, and packet loss\cite{HHH,openTM,OpenNetmon}. Two key limitations of rule-based counters emerged: 1) measurement data are tightly coupled to the flow-matching rules required to implement the forwarding policy; and 2) data collection led to high overhead on both controller(s) and switches (additional network traffic, CPU processing, and time).  Subsequent research mitigated these drawbacks using approximation algorithms and data structures (\ie sketches and bloom filters) to keep short summaries of traffic characteristics\cite{ OpenSketch, UnivMon}, with provable bounds on accuracy when assigned a certain amount of resources. But they had limited applicability and/or re-usability, as they were tailored to specific measurement problems that required special-purpose, hard-coded algorithms. 

Recent trends on programmable data planes\cite{P4} allow the definition and implementation of somewhat arbitrary measurement algorithms\cite{DAPPER,FlowRadar,LossRadar} that can be installed on programmable forwarding elements. However, current measurement approaches propose only tailored solutions to specific problems, without providing general and reusable abstractions to ease network measurements specification. %t is not clear how programmable data planes can foster re-usability for arbitrary run-time measurement needs.

In-band Network Telemetry \cite{INT}, a practical realization of TPP\cite{TPP}, aims at improving the visibility of the network operating conditions by gathering measurement data (\ie queuing, delay, losses and utilization) along the path taken by the packets. However, INT's applicability is limited by design and narrows its focus to the areas of performance monitoring and troubleshooting. Its techniques are not immediately applicable to other scenarios. Also, it cannot specify stateful measurement algorithms. Supporting explicit composability, \ourlang provides INT's benefits while also covering a wider range of measurements needs. 

Marple\cite{Marple} proposes high-level language abstractions to ease expressiveness of measurement tasks. It focuses on providing aggregation of linear-in-state query results directly in the switch, but is limited to performance-related measurements. Its technique relies on the ability to export state from switches on hash collisions towards an offline collector, where partial results are then aggregated. 
%\RM{functionality that could be also be beneficial for \ourlang stateful primitives.}

Sonata\cite{Sonata} proposes a data-streaming model for measurement queries with iterative refinement, modulating which data collection functions run in-network and which ones run at stream processors.
While Sonata and \ourlang share measurement flexibility as a goal, and both achieve it via composability, \ourlang goes beyond Sonata in the array of provided measurement primitives at switches. Sonata could exploit our primitives to deploy richer queries or optimize their execution plan based on resource requirements.

%Sonata\cite{Sonata} and Marple\cite{Marple} propose high-level language abstractions to ease the process of expressing measurement tasks. Marple focuses on providing aggregation of linear-in-state query results directly in the switch, but is limited to performance-related measurements. Its technique relies on the ability to export state from switches on hash collisions with offline aggregation, functionality that could be also be beneficial for \ourlang stateful primitives. Sonata proposes a data streaming model with iterative refinement and planning to drive in-network query execution and data collection. Our approach is complementary: we focus on switch-provided measurement capabilities. Sonata's runtime could exploit our primitives model to optimally answer queries by automatically configuring the best primitives to be used in each measurement interval. 

%Marple and Sonata focus on how to aggregate linear-in-state query results in a switch but are not concerned with providing the primitives required for measurements that span multiple switches (such as tagging). We see our contributions complementary to those of Marple and Sonata. In fact, these systems could exploit \ourlang primitives to optimally answer queries by automatically configuring the best primitives to be used in each measurement interval.

Finally, Allman et al.~\cite{Principles-for-Measurability} underline the need for built-in protocol support to aid measurements. They propose end-host and hop-by-hop support to gather measurement data along a packet's path. However, their approach is limited as it focuses on the host's view of the network. Also, they do not provide mechanisms to perform stateful processing of measurement data in the data plane. %and cannot efficiently address network-wide concerns (\eg traffic matrix, changes, heavy hitters).

\section{Conclusions and Future Work}
\label{sec:end}

We advocate that true software-defined measurement should start with  configurable and reusable \textit{primitives}---basic \textit{building-blocks} that, when properly composed, can support a wide variety of network measurements. That is, Measurements As FIrst-class Artifacts.

We proposed a set of orthogonal primitives that can be exported by switches to implement measurement tasks. We implemented a \ourlang prototype and showed how our principled approach
is capable of deriving concise and easy to understand specifications of measurement activity. We demonstrated, by using our primitives, how a wide range of measurement tasks can be easily solved using \ourlang. In future work, we plan to explore the necessary abstractions and protocols to deploy measurement tasks dynamically across a set of switches.
%As well, another possible direction for the work is the automatic verification of the correctness of \ourlang measurement programs. 

%\newpage

\vspace{2mm}\noindent\noindent {\footnotesize \textit{\textbf{Acknowledgments.}} This work was partially supported by Funda\c{c}\~{a}o para a Ci\^{e}ncia e Tecnologia (FCT) and Feder  projects with references PTDC/EEI-COM/29271/2017 (Cosmos) and UID/ CEC/ 50021/ 2019. Paolo Laffranchini was supported by a fellowship from the Erasmus Mundus Joint Doctorate in Distributed Computing (EMJD-DC) program funded by the European Commission (EACEA) (FPA 2012-0030).
We thank David Walker and Leonid Ryzhyk for their constructive comments on this work.}

\bibliographystyle{IEEEtran}
\bibliography{bibliography} 

\clearpage
\onecolumn
\appendix
\section{Implementation with primitives API}
\label{sec:appendix}

%\subsection*{Flow volume and duration\cite{openTM,OpenWatch,OpenNetmon,HHH,PayLess,FlowSense,DevoFlow}}
%\input{api-code/openflow-code.tex}

%%%%%%%%%%%%%%%%%%%%%%%%%%%%%%%%%%%%%%%%%%%%%%%%%%%%%%%%%%%%%%%%%%%%%%%%%%%%%%%%%%%%%%%%%%%%%%%%%%%%%%%%%%%%%%%%%%%% %%%%%%%%%%%%%%%%%%%%%%%%%%%%%%%%%%%%%%%%%%%%%%%%%%%%%%%%%%%%%%%%%%%%%%%%%%%%%%%%%%%%%%%%%%%%%%%%%%%%%%%%%%%%%%%%%%%% %%%%%%%%%%%%%%%%%%%%%%%%%%%%%%%%%%%%%%%%%%%%%%%%%%%%%%%%%%%%%%%%%%%%%%%%%%%%%%%%%%%%%%%%%%%%%%%%%%%%%%%%%%%%%%%%%%%% 
											% TABLE APPENDIX %
          									% TABLE APPENDIX %

\begin{table*}[t!]
\scriptsize
\begin{adjustwidth}{-0.25cm}{}
\begin{tabular}{|m{8em} m{3em} | m{50em} |}

\hline  
	\onecolmn{@{\extracolsep{\fill}}C}{8em}{\textbf{Measurement \newline use cases}} & 
    \onecolmn{C}{2em}{\textbf{Bib.}} &
    \onecolmn{C}{65em}{\textbf{Implementation with Primitives API}}

\\\hline 
	\textbf{\textit{Flow volume and duration}} &
    \cite{openTM,OpenWatch,OpenNetmon,HHH,PayLess,FlowSense,DevoFlow} & 
    \begin{lstlisting}[language=mafia]
flowid = Key(ip.src,ip.dest,tcp.src,tcp.dest,ip.proto)
now_ts = Timestamp();
byte_counter   = HashMap(key=flowid, size=1024, type=Counter(width=32));
packet_counter = HashMap(key=flowid, size=1024, type=Counter(width=32));
start_ts       = HashMap(key=flowid, size=1024, type=Timestamp());
flow_duration  = HashMap(key=flowid, size=1024, type=Counter(width=32));
   
pkts 
 ^$\seqlsym$^ ( byte_counter.set(byte_counter + pkt.size) ^$\parlsym$^ packet_counter.set( packet_counter + 1) )
    ^$\parlsym$^
    ( 
      ( match(start_ts == 0) ^$\seqlsym$^ timestamp(start_ts) )
      ^$\parlsym$^
      ( match(start_ts != 0)
        ^$\seqlsym$^ timestamp(now_ts)
        ^$\seqlsym$^ flow_duration.set(now_ts - start_ts)) )
    )

\end{lstlisting}

    %\onecolmn{@{\extracolsep{\fill}}C}{55em}{\input{api-code/openflow}}
    
\\\hline 
	\textbf{\textit{Approximate \newline flow volume}}\newline(Count-Min Sketch) & 
    \cite{DevoFlow,OpenSketch,Sonata,SCREAM,SketchCountMin} & 
    %%%%%%%%%%%%%%%%%%%%%%%%%%%%%%%%%%%%%%%%%% COUNTMIN-SKETCH %%%%%%%%%%%%%%%%%%%%%%%%%%%%%%%%%%%%%%%%%%%

%Sketch(^$\lambda(hash)$^:{countmin_sketch = countmin_sketch + 1})
\begin{lstlisting}[language=mafia]
flowid = Key(ip.src,ip.dest,tcp.src,tcp.dest,ip.proto)
flow_size = Sketch(alg="countmin", nhash=4, key=flowid, size=256, width=32); 

pkts ^$\seqlsym$^ flow_size.set(flow_size + pkt.size)
\end{lstlisting}

%%%%%%%%%%%%%%%%%%%%%%%%%%%%%%%%%%%%%%%%%% COUNTMIN-SKETCH %%%%%%%%%%%%%%%%%%%%%%%%%%%%%%%%%%%%%%%%%%%

%\\\hline
%    \textbf{\textit{Flow frequency}}\newline(Count-sketch) & 
%    \cite{SketchCount,UnivMon} & 
%    \input{api-code/sketch-count}
    
\\\hline
	\textbf{\textit{Flow cardinality}}\newline(PCSA Sketch) & 
    \cite{OpenSketch,SketchPCSA} & 
    %%%%%%%%%%%%%%%%%%%%%%%%%%%%%%%%%%%%%%%%%% PCSA %%%%%%%%%%%%%%%%%%%%%%%%%%%%%%%%%%%%%%%%%%%

%Sketch(^$\lambda_u$^:{1}, pcsa_sketch)
\begin{lstlisting}[language=mafia]
flowid = Key(ip.src,ip.dest,tcp.src,tcp.dest,ip.proto)
num_flows = Sketch(alg="pcsa",key=flowid,nhash=1,size=128); 

pkts ^$\seqlsym$^ num_flows.update()
\end{lstlisting}

%%%%%%%%%%%%%%%%%%%%%%%%%%%%%%%%%%%%%%%%%% PCSA %%%%%%%%%%%%%%%%%%%%%%%%%%%%%%%%%%%%%%%%%%%

\\\hline
	\textbf{\textit{Flow cardinality}}\newline(HyperLogLog Sketch) & 
    \cite{SCREAM,SketchHyperLogLog} & 
    %%%%%%%%%%%%%%%%%%%%%%%%%%%%%%%%%%%%%%%%%% HYPERLOGLOG %%%%%%%%%%%%%%%%%%%%%%%%%%%%%%%%%%%%%%%%%%%

%Sketch(^$\lambda(hash)$^:{hll_sketch = hll_sketch + count_zero_bits(key)}, hll_sketch)

\begin{lstlisting}[language=mafia]
flowid = Key(ip.src,ip.dest,tcp.src,tcp.dest,ip.proto)
num_flows = Sketch(alg="hyperloglog",key=flowid,nhash=1,size=256); 

pkts ^$\seqlsym$^ num_flows.update()
\end{lstlisting}

%%%%%%%%%%%%%%%%%%%%%%%%%%%%%%%%%%%%%%%%%% HYPERLOGLOG %%%%%%%%%%%%%%%%%%%%%%%%%%%%%%%%%%%%%%%%%%%

\\\hline 
	\textbf{\textit{Counter \newline thresholds}} & 
    \cite{DevoFlow} & 
    %%%%%%%%%%%%%%%%%%%%%%%%%%%%%%%%%%%%%%%%%% DEVOFLOW %%%%%%%%%%%%%%%%%%%%%%%%%%%%%%%%%%%%%%%%%%%

\begin{lstlisting}[language=mafia]
flowid = Key(ip.src,ip.dest,tcp.src,tcp.dest,ip.proto)
byte_counter   = HashMap(key=flowid, size=1024, type=Counter(width=32));
packet_counter = HashMap(key=flowid, size=1024, type=Counter(width=32));

pkts
  ^$\seqlsym$^ packet_counter.set(packet_counter + 1)
  ^$\seqlsym$^ match(packet_counter > ^\textsl{PACKET\_THRESHOLD}^)
  ^$\seqlsym$^ duplicate(pkts_exceeded)

pkts
  ^$\seqlsym$^ byte_counter.set(byte_counter + pkt.size)
  ^$\seqlsym$^ match(byte_counter > ^\textsl{BYTE\_THRESHOLD}^)
  ^$\seqlsym$^ duplicate(bytes_exceeded)

pkts_exceeded ^$\seqlsym$^ collect(COLLECTOR)

bytes_exceeded ^$\seqlsym$^ collect(^\textsl{COLLECTOR}^)
\end{lstlisting}

%%%%%%%%%%%%%%%%%%%%%%%%%%%%%%%%%%%%%%%%%% DEVOFLOW %%%%%%%%%%%%%%%%%%%%%%%%%%%%%%%%%%%%%%%%%%%

\\\hline
\end{tabular}
\end{adjustwidth}
  \vspace{2mm}
  \caption{Measurement use cases specification using our primitive-based API (1/4)}
  \label{tab:table-appendix-a}
\end{table*}
                                      
\begin{table*}[t!]
\scriptsize
\begin{adjustwidth}{-0.25cm}{}
\begin{tabular}{|m{8em} m{3em} | m{50em} |}

\hline  
	\onecolmn{@{\extracolsep{\fill}}C}{8em}{\textbf{Measurement \newline use cases}} & 
    \onecolmn{C}{3em}{\textbf{Bib.}} &
    \onecolmn{C}{65em}{\textbf{Implementation with Primitives API}}

\\\hline 
	\textbf{\textit{Stochastic \newline Sampling}} & 
    \cite{flexam,opensample,SampleAndPick} & 
    %%%%%%%%%%%%%%%%%%%%%%%%%%%%%%%%%%%%%%%%%% FLEXAM %%%%%%%%%%%%%%%%%%%%%%%%%%%%%%%%%%%%%%%%%%%

\begin{lstlisting}[language=mafia]
pkts ^\seqlsym^ match(random([0:100]) < ^\textsc{SamplingRatio}^) ^\seqlsym^ duplicate(samples) )

samples ^\seqlsym^  collect(^\textsl{COLLECTOR}^)
\end{lstlisting}

%%%%%%%%%%%%%%%%%%%%%%%%%%%%%%%%%%%%%%%%%% FLEXAM %%%%%%%%%%%%%%%%%%%%%%%%%%%%%%%%%%%%%%%%%%%

\\\hline 
	\textbf{\textit{Deterministic \newline Sampling}} & 
    \cite{flexam} & 
    %%%%%%%%%%%%%%%%%%%%%%%%%%%%%%%%%%%%%%%%%% FLEXAM %%%%%%%%%%%%%%%%%%%%%%%%%%%%%%%%%%%%%%%%%%%

\begin{lstlisting}[language=mafia]
flowid = Key(ip.src,ip.dest,tcp.src,tcp.dest,ip.proto)
n     = HashMap(key=flowid, size=1024, type=Counter(width=32))
m     = HashMap(key=flowid, size=1024, type=Counter(width=32))
delta = HashMap(key=flowid, size=1024, type=Counter(width=32))

pkts 
 ^$\seqlsym$^ (
      ( match(delta < ^\textsl{SKIP}^) ^$\seqlsym$^ delta.add(1) )
      ^$\parlsym$^
      ( match(delta >= ^\textsl{SKIP}^ && m < ^\textsl{NUM\_SAMPLES}^)
          ^$\seqlsym$^ m.set(m + 1))
          ^$\seqlsym$^ duplicate(samples) )
      ^$\parlsym$^
      (
        n.set(n + 1)
        ^$\seqlsym$^ match(n >= ^\textsl{NUM\_PACKETS}^)
        ^$\seqlsym$^ n.reset()
        ^$\seqlsym$^ m.reset()
        ^$\seqlsym$^ delta.reset()
      )
    )

samples ^$\seqlsym$^ collect(^\textsl{COLLECTOR}^)
\end{lstlisting}

%%%%%%%%%%%%%%%%%%%%%%%%%%%%%%%%%%%%%%%%%% FLEXAM %%%%%%%%%%%%%%%%%%%%%%%%%%%%%%%%%%%%%%%%%%%

\\\hline 
	\textbf{\textit{Postcard \newline generation}} & 
    \cite{netsight} & 
    %%%%%%%%%%%%%%%%%%%%%%%%%%%%%%%%%%%%%% NETSIGHT %%%%%%%%%%%%%%%%%%%%%%%%%%%%%%%%%%%%%%%

\begin{lstlisting}[language=mafia]
pkts
  ^$\seqlsym$^ duplicate(postcards)

postcards
  ^$\seqlsym$^ tag(ipv4.checksum, pkt.input_port) 
  ^$\seqlsym$^ tag(ipv4.identification, pkt.output_port) 
  ^$\seqlsym$^ tag(ipv4.tos, switchid) 
  ^$\seqlsym$^ collect(^\textsl{COLLECTOR}^)
\end{lstlisting}

%%%%%%%%%%%%%%%%%%%%%%%%%%%%%%%%%%%%%% NETSIGHT %%%%%%%%%%%%%%%%%%%%%%%%%%%%%%%%%%%%%%%

\\\hline
	\textbf{\textit{Trajectory \newline encoding}} & 
    \cite{VeriDP} & 
   	%%%%%%%%%%%%%%%%%%%%%%%%%%%%%%%%%%%%%%%%%% VERIDP %%%%%%%%%%%%%%%%%%%%%%%%%%%%%%%%%%%%%%%%%%%

\begin{lstlisting}[language=mafia]
//code executed at ingress switch
now = Timestamp() ;
flowid = key(ip.src,ip.dest,tcp.src,tcp.dest,ip.proto)
verify_time = HashMap(key=flowid, size=1024, type=Timestamp()) ;

pkts 
  ^$\seqlsym$^ timestamp(now)
  ^$\seqlsym$^ match( verify_time - now > ^\textsl{THRESHOLD}^)
  ^$\seqlsym$^ timestamp(verify_time)
  ^$\seqlsym$^ tag(ipv4.tos, switch.id)
  ^$\seqlsym$^ tag(ipv4.identification, pkt.input_port)
  ^$\seqlsym$^ tag(pkt.ipv4.tos, pkt.ipv4.tos|0x1)

//code executed at intermediate switch
location = key(pkt.input_port, switch.id, pkt.output_port)
trajectory = BloomFilter(alg="membership", nhash=4, key=location, size=16);

pkts
  ^$\seqlsym$^ match(pkt.ipv4.tos & 0x1 == 0x1)
  ^$\seqlsym$^ trajectory.insert()
  ^$\seqlsym$^ tag(ipv4.checksum, ipv4.checksum | trajectory.read()))
  ^$\seqlsym$^ trajectory.reset()

//code executed at egress switch
pkts
  ^$\seqlsym$^ match(pkt.ipv4.tos & 0x1 == 0x1)
  ^$\seqlsym$^ tag(ipv4.identification, pkt.output_port)
  ^$\seqlsym$^ tag(ipv4.tos, switch.id)
  ^$\seqlsym$^ duplicate(reports)

reports ^$\seqlsym$^ collect(^\textsl{COLLECTOR}^)
\end{lstlisting}

%%%%%%%%%%%%%%%%%%%%%%%%%%%%%%%%%%%%%%%%%% VERIDP %%%%%%%%%%%%%%%%%%%%%%%%%%%%%%%%%%%%%%%%%%%

\\\hline
\end{tabular}
\end{adjustwidth}
  \vspace{2mm}
  \caption{Measurement use cases specification using our primitive-based API (2/4)}
  \label{tab:table-appendix-b}
\end{table*}

\begin{table*}[t!]
\scriptsize
\begin{adjustwidth}{-0.25cm}{}
\begin{tabular}{|m{8em} m{3em} | m{50em} |}

\hline  
	\onecolmn{@{\extracolsep{\fill}}C}{8em}{\textbf{Measurement \newline use cases}} & 
    \onecolmn{C}{3em}{\textbf{Bib.}} &
    \onecolmn{C}{65em}{\textbf{Implementation with Primitives API}}

\\\hline 
	\textbf{\textit{Two-phase \newline heavy hitters}} &
     & 
    %%%%%%%%%%%%%%%%%%%%%%%%%%%%%%%%%%%%%%%%%% HEAVY HITTERS %%%%%%%%%%%%%%%%%%%%%%%%%%%%%%%%%%%%%%%%%%%

\begin{lstlisting}[language=mafia]

flowid = Key(ip.src,ip.dest,tcp.src,tcp.dest,ip.proto)
total = Counter(width=32)
nbytes = Sketch(alg="count-min",nhash=4,key=flowid,size=256,width=32)
hh = BloomFilter(alg="membership",key=flowid,nhash=4,size=64)
hh_bytes = HashMap(key=flowid,size=1024,type=Counter(width=32))

window(mment_interval)
          
pkts
  ^$\seqlsym$^ match(pkt.input_port == ^\textsl{PORT}^) 
  ^$\seqlsym$^ total.set(total + pkt.size)
  ^$\seqlsym$^ (
       ( match(!hh.test())
           ^$\seqlsym$^ nbytes.set(nbytes + pkt.size)
           ^$\seqlsym$^ match(nbytes.min() / total > ^\textsl{THRESHOLD}^) 
           ^$\seqlsym$^ hh.insert()
           ^$\seqlsym$^ hh_bytes.set(nbytes.min())
           ^$\seqlsym$^ duplicate(hh_alarms) )
       ^$\parlsym$^
        ( match(hh.test())
            ^$\seqlsym$^ hh_bytes.set(hh_bytes + pkt.size) )
     )

hh_alarms
  ^$\seqlsym$^ tag(ipv4.checksum, nbytes.min())
  ^$\seqlsym$^ collect(^\textsl{CONTROLLER}^)

// Control traffic to retrieve heavy hitters volume.
ctrl
  ^$\seqlsym$^ match(pkt.request == ^\textsl{HH\_VOLUME}^)
  ^$\seqlsym$^ duplicate(get_hh_volume)

get_hh_volume
  ^$\seqlsym$^ tag(pkt.hh_volume, hh_bytes)
  ^$\seqlsym$^ collect(^\textsl{CONTROLLER}^)

  
\end{lstlisting}

%%%%%%%%%%%%%%%%%%%%%%%%%%%%%%%%%%%%%%%%%% HEAVY HITTERS %%%%%%%%%%%%%%%%%%%%%%%%%%%%%%%%%%%%%%%%%%%

    %\onecolmn{@{\extracolsep{\fill}}C}{55em}{\input{api-code/openflow}}
    
\\\hline 
	\textbf{\textit{Top-k Congested Flows}}& 
     & 
    %%%%%%%%%%%%%%%%%%%%%%%%%%%%%%%%%%%%%%%%%% TOP K CONGESTED %%%%%%%%%%%%%%%%%%%%%%%%%%%%%%%%%%%%%%%%%%%

\begin{lstlisting}[language=mafia]
// Code executed at first hop:
pkts 
  ^$\seqlsym$^ tag(ipv4.tos, ipv4.tos | 0x1) 
  ^$\seqlsym$^ tag(ipv4.id, pkt.in_queue_length))

// Code executed at intermediate hops:
q_len = Counter(width=32);
pkts
  ^$\seqlsym$^ match(ipv4.tos & 0x1 == 0x1) 
  ^$\seqlsym$^ q_len.set(ipv4.id + pkt.in_queue_length)
  ^$\seqlsym$^ tag(ipv4.id, q_len)

// Code executed at last hop:
flowid = Key(ip.src,ip.dest,tcp.src,tcp.dest,ip.proto)
total_pkts = 
  Sketch(alg="countmin",key=flowid,nhash=4,size=1024,w=32)
path_q_len = 
  Sketch(alg="countmin",key=flowid,nhash=4,size=1024,w=32)

pkts.window(5s)
  ^$\seqlsym$^ match(ipv4.tos & 0x1 == 0x1)
  ^$\seqlsym$^ total_pkts.set(total_pkts + 1)
  ^$\seqlsym$^ path_q_len.set(path_q_len + ipv4.id)  
\end{lstlisting}

%%%%%%%%%%%%%%%%%%%%%%%%%%%%%%%%%%%%%%%%%% TOP K CONGESTED %%%%%%%%%%%%%%%%%%%%%%%%%%%%%%%%%%%%%%%%%%%

\\\hline 
	\textbf{\textit{Path Changes}} & 
     & 
    %%%%%%%%%%%%%%%%%%%%%%%%%%%%%%%%%%%%%%%%%% PATH CHANGES %%%%%%%%%%%%%%%%%%%%%%%%%%%%%%%%%%%%%%%%%%%

\begin{lstlisting}[language=mafia]
// Code to be executed at intermediate switches
location = Key(pkt.input_port, switch.id, pkt.output_port)
location_bf = BloomFilter(
          alg="membership",key=location,nhash=4,size=32)

pkts  
  ^$\seqlsym$^ location_bf.init(ipv4.checksum)
  ^$\seqlsym$^ trajectory.set()
  ^$\seqlsym$^ tag(ipv4.checksum, ipv4.checksum | location_bf)
  ^$\seqlsym$^ location_bf.reset()
  
// Code to be executed at the packet's last hop
flowid = Key(ip.src,ip.dest,tcp.src,tcp.dest,ip.proto)
paths_sketch = Sketch(
    alg="store",key=flowid,nhash=4,size=256,width=16)
n_change_sketch = Sketch(
    alg="countmin",key=flowid,nhash=4,key=flowid,size=256)

pkts.window(10 ^$\times$^ ^\textsl{RTT}^)
  ^$\seqlsym$^  match(!paths_sketch.any(ipv4.checksum))
  ^$\seqlsym$^  paths_sketch.set(ipv4.checksum)
  ^$\seqlsym$^  n_change_sketch.set(n_change_sketch + 1)
\end{lstlisting}

%%%%%%%%%%%%%%%%%%%%%%%%%%%%%%%%%%%%%%%%%% PATH CHANGES %%%%%%%%%%%%%%%%%%%%%%%%%%%%%%%%%%%%%%%%%%%

\\\hline

\end{tabular}
\end{adjustwidth}
  \vspace{2mm}
  \caption{Measurement use cases specification using our primitive-based API (3/4)}
  \label{tab:table-appendix-c}
\end{table*}

\begin{table*}[t!]
\scriptsize
\begin{adjustwidth}{-0.25cm}{}
\begin{tabular}{|m{8em} m{3em} | m{50em} |}
    
\hline  
	\onecolmn{@{\extracolsep{\fill}}C}{8em}{\textbf{Measurement \newline use cases}} & 
    \onecolmn{C}{3em}{\textbf{Bib.}} &
    \onecolmn{C}{65em}{\textbf{Implementation with Primitives API}}
    
\\\hline
	\textbf{\textit{Path Change Latency}}& 
     & 
    %%%%%%%%%%%%%%%%%%%%%%%%%%%%%%%%%%%%%%%%%% PATH CHANGES LATENCY %%%%%%%%%%%%%%%%%%%%%%%%%%%%%%%%%%%%%%%%%%%

\begin{lstlisting}[language=mafia]
// Code to be executed on all switches updating rules
change_ts = Timestamp();
l_clock = Counter(width=8);
pkts
  ^$\seqlsym$^ match(segway_header.msg == GoodToMove) 
  ^$\seqlsym$^ l_clock.set(max(l_clock + 1, segway_header.ts))
  ^$\seqlsym$^ tag(segway_header.ts, l_clock)
  ^$\seqlsym$^ duplicate(end_of_update) 

end_of_update
  ^$\seqlsym$^ timestamp(change_ts) 
  ^$\seqlsym$^ tag(segway_header.time, change_ts)
  ^$\seqlsym$^ tag(segway_header.ts, l_clock)
  ^$\seqlsym$^ collect(^\textsl{SEGWAY\_CONTROLLER}^)
\end{lstlisting}

%%%%%%%%%%%%%%%%%%%%%%%%%%%%%%%%%%%%%%%%%% PATH CHANGES LATENCY %%%%%%%%%%%%%%%%%%%%%%%%%%%%%%%%%%%%%%%%%%%

\\\hline
\end{tabular}

\end{adjustwidth}
  \vspace{2mm}
  \caption{Measurement use cases specification using our primitive-based API (4/4)}
  \label{tab:table-appendix-d}
\end{table*}

											% TABLE APPENDIX %
%%%%%%%%%%%%%%%%%%%%%%%%%%%%%%%%%%%%%%%%%%%%%%%%%%%%%%%%%%%%%%%%%%%%%%%%%%%%%%%%%%%%%%%%%%%%%%%%%%%%%%%%%%%%%%%%%%%% %%%%%%%%%%%%%%%%%%%%%%%%%%%%%%%%%%%%%%%%%%%%%%%%%%%%%%%%%%%%%%%%%%%%%%%%%%%%%%%%%%%%%%%%%%%%%%%%%%%%%%%%%%%%%%%%%%%% %%%%%%%%%%%%%%%%%%%%%%%%%%%%%%%%%%%%%%%%%%%%%%%%%%%%%%%%%%%%%%%%%%%%%%%%%%%%%%%%%%%%%%%%%%%%%%%%%%%%%%%%%%%%%%%%%%%% 

\end{document}